\documentstyle[preprint,epsf,aps]{revtex}
\begin{document}
\draft
\title
{Quadrupolar and magnetic ordering in CeB$_6$
}
\author{Gennadi Uimin\cite{gu}} 
\address{
Institut f\"ur Theoretische Physik, Universit\"at zu K\"oln,
D-50937 K\"oln, Germany\\
and\\
DRFMC, SPSMS/MDN, CENG, 17, rue des Martyrs, 38054 Grenoble Cedex 9, France}
\maketitle

\begin{abstract}
The quadrupolar ordering in CeB$_6$ is explained in terms of the
electrostatic interaction of quadrupolar moments arranged into a
simple cubic lattice.  The representation of magnetic and quadrupolar
moments by means of quasispins of two kinds is employed.  A linear
increase of the quadrupolar transition temperature $T_Q(H)$ with
applied magnetic field and its further re-entrance are described
using a generalized spherical model which is well adjusted to a
particular problem of the quadrupolar ordering in CeB$_6$.
The theory naturally explains the growing specific heat jump at
$T_Q(H)$ with increasing magnetic field.  The role of the quadrupolar
ordering in the formation of the magnetic ordering, as well as the
possible critical experiments and applications to other rare-earth
compounds, are discussed.
\\
\\
\\
Keywords: Quadrupolar ordering
\end{abstract}
\pacs{71.70.-d, 75.10.-b, 75.50.-y}
\section{Introduction}
The aim of the present paper is to discuss the nature of the
quadrupolar ordering in CeB$_6$.  This compound is classified as a
dense Kondo system.  With decreasing temperature, the resistivity
grows logarithmically, attaining its maximum at $T\approx 3.2$ K
\cite{winzer}.  The Kondo temperature was initially estimated as
$T_K\approx 8$ K \cite{takase}.  Later this value was significantly
revised to a value of $T_K\approx 1$ K from the experimental data on
the magnetic susceptibility versus temperature \cite{sato84}.  This
revision was caused by an unusual picture of the crystal field
splitting, revealed in the Raman and neutron spectroscopic
measurements \cite{zirn}.
It is well-known that the crystalline electric field (CEF) of cubic
symmetry (the elementary cell, containing the Ce ion with its boron
environment, is shown in Fig. 1) results in splitting of the
Ce$^{3+}$ multiplet ($4f^1, {\cal J}\!=\!5/2, S\!=\!1/2, L\!=\!3$)
into a $\Gamma_7$ doublet and a $\Gamma_8$ quartet.  The ground state
of Ce$^{3+}$ in CeB$_6$ is realized as the well isolated $\Gamma_8$
quartet, and the $\Gamma_8-\Gamma_7$ CEF gap has been determined as 47
meV \cite{zirn}.  Prior to the results of Ref. \cite{zirn} many
difficulties in interpretation of experimental data had arisen in
connection with incorrect assumptions on the multiplet splitting:
The ground state level had generally been ascribed to $\Gamma_7$ (cf.,
however, \cite{burlet,ohkawa}).  The quadrupolar and magnetic
transitions were proved in the specific heat measurements
\cite{lee,fujita,peysson}, NMR \cite{kawa,takigawa} and neutron
diffraction \cite{horn,effantin} studies.  Since the typical ordering
temperatures (quadrupolar, $T_Q\approx 3.3$ K, and magnetic,
$T_N\approx 2.4$ K) are much smaller than the CEF splitting, for low
energy phenomena with $T$ not exceeding several tens Kelvin, it should
be legitimate to neglect a $\Gamma_7$ contribution and to deal with
$\Gamma_8$ only.

The quadrupolar ordering is characterized by the following features:
\begin{itemize}
\item
There are two lines in the $T\!-\!H$ phase diagram, which separate the
anti-ferro-quadrupolar (AFQ) phase from the complex antiferro-magnetic
(AFM) phases (see Fig. 2) and from the disordered (D)
phase.  The AFQ-D transition line, $T_Q(H)$, exhibits a highly
anisotropic behavior.  Starting at $T_Q\approx 3.3$ K, $T_Q$ increases
with $H$ at not very high magnetic field, and increases linearly,
$dT_Q(H)/dH>0.$ The re-entrant behavior of $T_Q(H)$, predicted
theoretically in \cite{uim1}, has not yet been confirmed
experimentally up to magnetic field of 18 Tesla
\cite{effantin,erk,lacerda} (for recent experimental data, see
Fig. 3).  It is worth noting, that the estimates from {\it
below} for the values of the critical field at the re-entrance and
$T_Q=0$ are 18 Tesla and 60 Tesla, respectively, according to
Ref. \cite{uim1}.
\item
There are contradictory AFQ patterns obtained in different microscopic
measurements: neutron \cite{effantin,erk}, NMR \cite{takigawa} and
$\mu$SR \cite{feyer}.  The interpretation of neutron experiments is
consistent with the AFQ patterns of the ${\bf Q}=[\frac 12 \frac 12
\frac 12]$ modulation, whereas the NMR and $\mu$SR measurements
display more complicated AFQ structures.  Note, that until now the
above-mentioned microscopic methods, {\it i.e.}, $^{11}$B--NMR,
neutron diffraction, as well as $\mu$SR, are used in non-zero magnetic
fields which generate magnetically ordered states.  Indeed, picking
up an identical modulation with the AFQ state, a magnetic ordering is
a secondary effect with respect to the primary quadrupolar ordering.
The theoretical approach developed in Ref. \cite{uim1} selects the
$[\frac 12 \frac 12 \frac 12]$--structure as energetically
preferential.
\item
The specific heat jump at $T_Q$ appears to be of order of magnitude
smaller than its counterpart at $T_N$ ($H=0$) \cite{lee,peysson}.
This points out an important role of fluctuations at the D--AFQ
transition.  This circumstance has been taken into account in
\cite{uim1} by employing the spherical model description of the
effective spin Hamiltonian.  The specific heat jump on the D--AFQ
transition line grows with $H$ \cite{peysson}.
\end{itemize}
The main features of the magnetic ordering have been presented in
\cite{effantin}:
\begin{itemize}
\item
With a magnetic field applied along $[1 1 1]$, the AFM-structure is
characterized by the wave-vector, either ${\bf k}_1=[\frac 14 \frac 14
\frac 12]$ or ${\bf k}_2=[\frac 14 \frac {\overline 1}4 \frac 12]$
(the single-{\bf k} structure at sufficiently high magnetic field), by
a couple of ${\bf k}$'s, ${\bf k}_1$ and ${\bf k}_2$ (the doublle-{\bf
k} structure in moderate fields), and by a mixture of differently
oriented domains at weak magnetic fields (see Fig. 2,
where all these magnetic phases are sketched out).
\item
The Bragg peaks at the wave-vectors ${\bf k}_1$ and/or ${\bf k}_2$ are
accompanied by $\overline{{\bf k}_1}=[\frac 14 \frac 14 0]$ and/or
$\overline{{\bf k}_2}=[\frac 14 \frac {\overline 1}4 0]$, respectively
\cite{effantin}.  Their occurrence is a sign of a crucial role of the
AFQ modulation, ${\bf Q}= [\frac 12 \frac 12 \frac 12]$, in the
formation of magnetic structures.  A possible double-{\bf k}
structure, identified in \cite{effantin}, is shown in
Fig. 4.
\end{itemize}

In the next Section an effective "separation" of spin and orbital
degrees of freedom is carried out by introducing a formalism,
according to which magnetic and quadrupolar moments can be properly
described by means of two Pauli matrices, $\bbox \sigma$ and
$\bbox\tau$.  Section III concerns the analysis of the AFQ
ground state and relevant excitations.  In Section IV, on the basis of
the two relevant interactions, Zeeman and quadrupolar, and using the
spherical model for picking up these interactions, we are able to
determine the shape of $T_Q(H)$. It occurs to be strongly anisotropic
in the $T\!-\!H$ plane.  Despite a perfect cubic lattice symmetry,
such a strong anisotropy is due to a spacial anisotropy of the
quadrupolar interaction.  Its conventional form, following from the
Coulomb's interaction, gives rise to a very soft mode of
$\tau$-excitations in the particular case of a simple cubic lattice.
Experimentally, strong fluctuations are indicated by a small specific
heat jump at the D-AFQ transition.  This is a reason for employing the
spherical model which is an appropriate tool for describing systems
with developed fluctuations.

The spherical model is applied for deriving analytical formulae for
the specific heat near the AFQ-D transition.  We also outline how the
magnetically ordered state can be generated by the quadrupolar
interaction via quantum fluctuations of orbital-like "spins",
$\tau${\footnotesize s}.  In Section V the $\sigma-\tau$
representation is used for the case of a single $f$ hole
(configuration $f^{13}$), which is likely ascribed to the rare-earth
compound TmTe.  In the concluding Section we discuss what kind of
experiments could be critical for establishing the nature of the AFQ
order unambigiously.
\section{Theoretical prerequisite}
\subsection{Representation of moments through the Pauli matrices.
Zeeman interaction}
We represent the set of the $\Gamma_8$ states with use of the $|{\cal
J}_z\rangle$ (abbreviation for $|L,S,{\cal J},{\cal J}_z\rangle$)
basis in the following form:
\begin{equation}
\label{j}
\psi_{1,\pm}=\sqrt{\frac 56}|\pm 5/2\rangle+\sqrt{\frac 16}|\mp 3/2\rangle,\quad
\psi_{2,\pm}=|\pm 1/2\rangle\,.
\end{equation}
\vspace{-5mm}

\noindent
The quartet constituents in Eq.(\ref{j}) are labelled in such a way
in order to make use of the Pauli matrices, $\bbox\sigma$ and
$\bbox\tau$, convenient.  For each $\ell$, the Kramers
doublet $\psi_{\ell,\pm}$ is defined as
\begin{equation}
\sigma_z\psi_{\ell,\pm}=\pm\frac 12\psi_{\ell,\pm}\,,\quad\sigma_+\psi_{\ell,-}
=\psi_{\ell,+}\;\; (\sigma_-\psi_{\ell,+}=\psi_{\ell,-})\,. 
\end{equation}
\vspace{-5mm}

\noindent
The orbital doublet $\psi_{1,\sigma}$ and $\psi_{2,\sigma}$ can be
suitably defined with using the pseudo-spin operator
$\bbox\tau$ as
\begin{eqnarray}
\tau_z\psi_{1,\sigma}=\frac 12\psi_{1,\sigma}\,,\; \tau_z\psi_{2,\sigma}=
-\frac 12\psi_{2,\sigma}\,, 
\quad
\tau_+\psi_{2,\sigma}=\psi_{1,\sigma}\;\;
(\tau_-\psi_{1,\sigma}=\psi_{2,\sigma})\,.  
\end{eqnarray}
\vspace{-5mm}

\noindent
This representation was proposed in \cite{ohkawa}; however,
expressions for the magnetic moment in terms of
$\bbox \sigma$ and $\bbox \tau$, given in
\cite{ohkawa}, are oversimplified.

In order to derive formulae for the moments ({\bf J}, {\bf S}, {\bf L},
{\bf M}), we need to calculate the matrix elements of, say, {\bf J}
over the set $\{\psi_{\ell,\sigma}\}$: 
\begin{eqnarray}
\label{j1}
<\!\psi_{1,\pm}|{\cal J}_z|\psi_{1,\pm}\!>=\pm \frac{11}6, 
<\!\psi_{2,\pm}|{\cal J}_z|\psi_{2,\pm}\!>=\pm \frac 12, \nonumber\\
<\!\psi_{2,\mp}|{\cal J}_\pm|\psi_{1,\pm}\!>=\frac 2{\sqrt{3}}, 
<\!\psi_{1,\mp}|{\cal J}_\pm|\psi_{2,\pm}\!>=\frac 2{\sqrt{3}}, \\
<\!\psi_{1,\pm}|{\cal J}_\pm|\psi_{1,\mp}\!>=\frac 53,
<\!\psi_{2,\pm}|{\cal J}_\pm|\psi_{2,\mp}\!>=3.\nonumber
\end{eqnarray}
Within the Russell-Saunders scheme, the matrix elements of the moments 
can be obtained from their ${\cal J}$ counterpart in accordance with 
$g$-factors of the $f^1$ multiplet:
\begin{eqnarray}
<\!..S..\!>=-\frac 17<\!..{\cal J}..\!>,\quad 
<\!..L..\!>=\frac 87<\!..{\cal J}..\!>, \quad 
<\!..M..\!>=\frac 67\mu_B<\!..{\cal J}..\!>.
\nonumber
\end{eqnarray}
\vspace{-3mm}

\noindent
Using the matrix elements (\ref{j1}) we can express the operator of magnetic 
moment {\bf M}, which is associated with $\Gamma_8$ as follows:
\begin{eqnarray}
\label{mom}
M_i=2\mu_B\sigma_i(1+\frac 87 T_i)\,,\quad i=x,y,z,
\end{eqnarray}
where
\begin{eqnarray}
\label{mom1}
T_z\!=\!\tau_z,\quad T_x\!=\!-\frac 12\tau_z\!+\!\frac{\sqrt{3}}2\tau_x,\quad
T_y\!=\!-\frac 12\tau_z\!-\!\frac{\sqrt{3}}2\tau_x.
\end{eqnarray}
The derivation of formulae (\ref{mom}) from the set of matrix elements
(\ref{j1}) is outlined in Appendix A.  Note, that the
$\tau_y$-component is not involved in (\ref{mom}).  For the Zeeman
interaction
\begin{equation}
\label{hz}
{\cal H}_Z=-H_i\sum_{\bf r}M_i({\bf r}),
\end{equation}
we shall use representation (\ref{mom})-(\ref{mom1}).  In (\ref{hz})
the sum runs over the Ce lattice sites. As usual, summation over
repeated indices ($i$, Cartesian coordinates) is supposed.

Let us discuss some simple properties of Hamiltonian (\ref{hz}), which are 
important in experimental applications to CeB$_6$.
\begin{enumerate}
\item
If the "orbital", {\it i.e.}, $\tau$-subsystem exhibits some AFQ order
characterized by the modulation vector {\bf Q}, then at $H\neq 0$ the
effective Zeeman term, acting on $\sigma${\footnotesize s}, produces,
{\it first}, the uniform $\sigma$-component, and {\it second}, the {\bf
Q}-modulated $\sigma$-components. Both are absent in zero field.  As a
result, a {\it uniform} magnetic field causes the {\bf Q}-modulated
magnetization.  This property of the AFQ phase has been used in
neutron, NMR and $\mu$SR experiments.  In the weak-field region, with
$H$ not exceeding few Tesla, the {\bf Q}-modulated magnetization is
linear in $H$.
\item
In the magnetically ordered phase (see Fig. 2) the
Bragg peaks are related either to the single {\bf k}-structure (either
${\bf k}_1=[\frac 14\frac 14\frac 12]$, or ${\bf k}_2=[\frac
14\frac{\overline 1}4\frac 12]$), or to the double {\bf k}-structure
(${\bf k}_1$ and ${\bf k}_2$).  These peaks at ${\bf k}_1$ and ${\bf
k}_2$ are accompanied by the Bragg peaks at $\overline{{\bf
k}}_1=[\frac 14\frac 140]$ and $\overline{{\bf k}}_2=[\frac
14\overline{\frac 14}0]$.  This fact can be easily understood if we
note, that magnetization (\ref{mom}) is related not only to the
$\sigma$-modulations (wave vectors ${\bf k}_1$ and/or ${\bf k}_2$),
but also to the ($\sigma\cdot\tau$)-modulations.  The latter
correspond to wave vectors ${\bf k}_1+{\bf Q}$ and/or ${\bf k}_2+{\bf
Q}$ with ${\bf Q}=[\frac 12\frac 12\frac 12]$.
\item
Non-interacting $\Gamma_8$ ionic states can be realized practically,
say, in La$_{1-x}$Ce$_x$B$_6$.  Owing to a non-trivial form of the
Zeeman interaction, magnetization is not aligned with ${\bf H}$ except
for a few special $H$-orientations, {\it e.g.}, [001], [110], [111],
and their equivalents.  At fixed $H$ the bigger energy gain is
achieved for directions of the [001]-type. This kind of {\bf
H}--anisotropy is an inherent property of the well-isolated $\Gamma_8$
states.
\end{enumerate}
\subsection{Quadrupolar interaction}
Not only the vector moments, but the quadrupolar moment $Q_{ij}$ 
($i,j\!=\!x,y,z$) on a Ce site as well, can be expressed in terms of the 
$\bbox\sigma$ and $\bbox\tau$ operators.  
For calculating the matrix elements of $Q_{ij}$ over the 
set $\{\psi_{\ell,\sigma}\}$,
\begin{eqnarray}
\label{qdef1}
<\!\psi_{\ell,\sigma}|Q_{ij}|\psi_{\ell ',\sigma '}\!\!>=\!
e\!\int\! d^3r\psi_{\ell,\sigma}^*({\bf r})\psi_{\ell ',\sigma '}({\bf r})
(3x_i x_j\!-\!\delta_{ij}r^2),
\nonumber
\end{eqnarray}
we can employ the Wigner-Eckhart theorem, according to which these matrix 
elements are proportional to the operator equivalents:
\begin{eqnarray}
\label{qdef}
<\!..|Q_{ij}|..\!\!>\,\propto\,
<\!..|\frac 12({\cal J}_i{\cal J}_j+{\cal J}_j{\cal J}_i)-
\frac 13\delta_{ij}
{\bf J}^2
|..\!\!>.
\nonumber
\end{eqnarray}
Given below are the matrix elements of the quadrupolar moment; we measure
them in units of
$$
Q_0=<\!\psi_{1,\sigma}|Q_{zz}|\psi_{1,\sigma}\!>,
$$
{\it i.e.}, 
\begin{equation}
\label{unit}
Q_0=e\int d^3r\psi_{1,\sigma}^*({\bf r})\psi_{1,\sigma}({\bf r}) 
(3z^2\!-\!r^2)\,.
\end{equation}

The $Q$ matrix elements can be classified as $\sigma$-independent
\begin{eqnarray}
\label{q_ind}
\displaystyle{
\begin{array}{l}
<\!\psi_{1,\sigma}|Q_{zz}|\psi_{1,\sigma}\!>=\!-\!
<\!\psi_{2,\sigma}|Q_{zz}|\psi_{2,\sigma}\!>=1, \\
<\!\psi_{1,\sigma}|Q_{xx}|\psi_{1,\sigma}\!>=
<\!\psi_{1,\sigma}|Q_{yy}|\psi_{1,\sigma}\!>=\!-\frac 12, \\
<\!\psi_{2,\sigma}|Q_{xx}|\psi_{2,\sigma}\!>=
<\!\psi_{2,\sigma}|Q_{yy}|\psi_{2,\sigma}\!>=\!\frac 12, \\
<\!\psi_{2,\sigma}|Q_{xx}|\psi_{1,\sigma}\!>=
-<\!\psi_{1,\sigma}|Q_{yy}|\psi_{2,\sigma}\!>=\!\frac{\sqrt{3}}2 ,
\end{array}}
\end{eqnarray}
as well as the $\sigma$-dependent 
\begin{eqnarray}
\label{q_dep}
\displaystyle{
\begin{array}{l}
<\!\psi_{2,+}|Q_{xy}|\psi_{1,+}\!>=
<\!\psi_{1,-}|Q_{xy}|\psi_{2,-}\!>=
\!i\frac{\sqrt{3}}8, \\
<\!\psi_{2,-}|Q_{yz}|\psi_{1,+}\!>=
<\!\psi_{2,+}|Q_{yz}|\psi_{1,-}\!>=
\!i\frac{\sqrt{3}}8, \\
<\!\psi_{2,+}|Q_{zx}|\psi_{1,-}\!>=-
<\!\psi_{1,+}|Q_{zx}|\psi_{2,-}\!>=\!\frac{\sqrt{3}}8 .
\end{array}}
\end{eqnarray}
We omit the Hermitian conjugated matrix elements in 
Eqs.(\ref{q_ind})-(\ref{q_dep}).

The matrix $\|Q\|$ can be written in the operator form as (see Appendix A for 
elementary explanations):
\begin{eqnarray}
\label{q_matr}
\|Q\|=Q_0\left(
\begin{array}{ccc}
2T_x & \mu _z & \mu _y \\
\mu _z & 2T_y & \mu _x \\  
\mu _y & \mu _x & 2T_z
\end{array}
\right) , 
\end{eqnarray}
where $\mu_i = \frac{\sqrt{3}}2\tau_y\sigma_i$.
The fact that $Q_{ij}$ contains the $\sigma$-variables signals that 
the quadrupolar interaction can be responsible not only for pure orbital 
interactions, but also for magnetic interactions.

The dependences of $M_i$ and $Q_{ij}$ on $\bbox \sigma$ and 
$\bbox\tau$ determine the time-reversal properties of the 
$\bbox\sigma$ and $\bbox\tau$ components. 
It is evident from (\ref{mom}) that 
$\bbox\sigma$$\,\rightarrow\,-\,$$\bbox\sigma$ under 
$t\rightarrow -\,t$, whereas $\tau_x$ and $\tau_z$ are unchanged. 
The off-diagonal components of $Q_{ij}$ require 
$\tau_y \rightarrow -\,\tau_y$ under the time-reversal transformation.
 
We suppose that the predominant contribution to the interactions of Ce
ions in CeB$_6$ comes from their quadrupolar interaction, the role of
which in Ce compounds was first mentioned by Bleaney \cite{bl} (for a
discussion on various forms of the quadrupolar interaction see, for
instance, \cite{birg,mor}).  We accept the form of the quadrupolar
interaction of the electrostatic origin, which is free of any model
assumptions.  Thus, our consideration is confined to the Zeeman and
quadrupolar interactions:
\begin{eqnarray}
{\cal H}={\cal H}_{\rm qd} +{\cal H}_{\rm Z}, \nonumber\\
{\cal H}_{\rm qd}={\displaystyle \sum_{{\bf r}\neq{\bf r}'}\sum_{i\dots n}}
{\cal A}_{ij,mn}({\bf r}\!-\!{\bf r}') 
Q_{ij}({\bf r})Q_{mn}({\bf r}'),
\label{qd}
\end{eqnarray}
where ${\cal A}_{ij,mn}({\bf r}\!-\!{\bf r}')$ is determined 
by the interaction $V_{q}({\bf r}-{\bf r}')$ 
of two quadrupolar moments located at ${\bf r}$ and ${\bf r}'$.  
The latter is given by
\begin{eqnarray}
\label{v}
V_{q}({\bf r})\!=\!\!{\displaystyle\frac 1{12 r^5}} 
\{2Q_{ij}(0)Q_{ij}({\bf r})\!-\!
20\,Q_{ij}(0)Q_{im}({\bf r})n_j n_m\nonumber \\
+35\,Q_{ij}(0)Q_{mn}({\bf r})n_i n_j n_m n_n)\},\;n_i\!=\!x_i/r\,.
\end{eqnarray}

\noindent
Thus, we obtain from (\ref{v})
\begin{eqnarray}
\label{a}
{\cal A}_{ij,mn}({\bf r})
\!=\!\frac 1{24 r^5}
\{(\delta_{im}\delta_{jn}\!+\!\delta_{in}\delta_{jm})+
35n_i n_j n_m n_n\nonumber\\-
5(\delta_{im}n_j n_n+\delta_{in}n_j n_m
+\delta_{jm}n_i n_n+\delta_{jn}n_i n_m)\}\,
\end{eqnarray}

\noindent
Evident are the following properties of ${\cal A}_{ij,mn}$'s with respect 
to permutation of indices:
$$
{\cal A}_{ij,mn}={\cal A}_{ji,mn}={\cal A}_{ij,nm}={\cal A}_{mn,ij}.
$$
The diagonal elements of matrix $\|Q\|$ give rise to the order parameter 
which is transformed according to representation $\Gamma_3$
characterized by two components ($\tau_x, \tau_z$) of $\bbox\tau$,
while the off-diagonal elements are related to symmetry $\Gamma_5$ and 
are characterized by the vector $\bbox\mu$. 
Keeping this in mind, we can rewrite ${\cal H}_{\rm qd}$ as follows:
\begin{eqnarray}
\label{qd1}
{\cal H}_{\rm qd}=\sum_{{\bf r}\neq{\bf r}'}\left[ 
{\cal A}_{\alpha\beta}({\bf r}\!-\!{\bf r}')\tau_\alpha({\bf r})
\tau_\beta({\bf r}')+{\cal B}_{ij}({\bf r}\!-\!{\bf r}')\mu_i({\bf r})
\mu_j({\bf r}')\right]\,,
\end{eqnarray}
where the Greek indices $\alpha,\beta$ prescribe summation over $x$-
and $z$-components only.  The expressions for ${\cal A}_{\alpha\beta}$
and ${\cal B}_{ij}$ are given in Appendix B. The Hamiltonian in
Eq.(\ref{qd1}) represents an evident separation of the orbital-like
and spin-like parts.

The magnetic exchange interactions are not relevant for a theoretical
analysis of the AFQ ordering in CeB$_6$.  This applies to a major part
of the phase diagram outside its low-temperature-and-weak-field part.
The latter requires the RKKY- and Kondo-like interactions to be
included.
\section{Towards quadrupolar ordering}
Taking the Fourier transform of the Hamiltonian (\ref{qd1}), we arrive
at its {\bf k}-diagonal form:
\begin{eqnarray}
\label{qdk}
{\cal H}_{\rm qd}=\sum_{\bf k}\left\{{\cal A}^{\alpha\beta}[{\bf k}]
\tau_{{\bf k},\alpha}^*\tau_{{\bf k},\beta}+{\cal B}^{ij}[{\bf k}]
\mu_{{\bf k},i}^*\mu_{{\bf k},j}\right\}\,.
\end{eqnarray}
We use the notation ${\cal A}^{\alpha\beta}[{\bf k}]$ and ${\cal
B}^{ij}[{\bf k}]$ for the Fourier transformed coupling constants at
general ${\bf k}$.  For high symmetry points of reciprocal space, such
as [000], $[\frac 12\frac 12\frac 12]$, $[\frac 12\frac 120]$ and
$[00\frac 12]$, as well as along the cubic edge $[\frac 12\frac
12\kappa]$, the Fourier transformed Hamiltonian (\ref{qdk}) becomes
completely diagonal:
\begin{eqnarray}
\label{qdkd} 
{\cal H}_{\rm qd}=\sum_{\bf k}\left\{{\cal A}^x_{\bf k} 
|\tau_{{\bf k},x}|^2+{\cal A}^z_{\bf k} 
|\tau_{{\bf k},z}|^2\right.
\left.+{\cal B}^x_{\bf k}( |\mu_{{\bf k},x}|^2+
|\mu_{{\bf k},y}|^2)+{\cal B}^z_{\bf k}|\mu_{{\bf k},z}|^2
\right\}\,.
\end{eqnarray}
Table I shows the result of numerical calculations for the
coefficients in units of $Q_0^2/a^5$ where $a$ denotes the lattice
constant.  Let us estimate the order of such an energy unit.  In doing
so, we return to definition (\ref{unit}), and then, performing the
radial and angular integrations, we get:
$$
Q_0=-\frac {16}{35}e<\!r_{\rm f}^2\!>\,.
$$
Then, the energy unit becomes 
\begin{eqnarray}
\left(\frac{16}{35}\frac{<\!r_{\rm f}\!>}a\right)^2\frac{e^2}a.
\end{eqnarray}
  For CeB$_6$, the lattice constant
$a\approx$ 4 \AA, the $f$-electron radius $r_{\rm f}\approx 0.4$ \AA,
the lattice Coulomb unit $e^2/a\approx$ 3 eV, and we arrive at the
$Q_0$-unit of order 1 K (cf., however, \cite{mor_2}).  From Table I
one can see that the coefficients ${\cal B}_{{\bf k}}^\alpha$ are
small as compared to the dominant ones, ${\cal A}${\footnotesize s}.
Additional smallness of the ${\cal B}$-terms comes from the fact that
the maximal value of $\mu_i^2$ is 16/3 times smaller than the maximal
value of $\tau_\alpha^2$.  Thus, it seems appropriate to simplify the
model by neglecting the ${\cal B}$-terms and to employ the simplified
version of ${\cal H}_{\rm qd}$ in its purely orbital $\tau$-form:
\begin{equation}
\label{orb}
{\cal H}_{\rm orb} = \sum_{{\bf r}\neq{\bf r}'}\sum_{\alpha\beta} 
{\cal A}_{\alpha\beta}({\bf r}\!-\!{\bf r}')
\tau_{\alpha}({\bf r})\tau_{\beta}({\bf r}')\,.
\end{equation}
 
According to Table I, the global energy minimum could be achieved at
${\bf k}= {\bf Q}$.  Not only the high-symmetry points of reciprocal
space, but also wave vectors of a general position have been checked
numerically in order to identify ${\bf Q}$ with the global energy
minimum.  It is necessary to emphasize that the energies at ${\bf
k}=[\frac 12\frac 120]$ and ${\bf Q}$ are only slightly different.
This is an indication of pronounced soft modes along the directions of
cubic edges, {\it i.e.}, $[\frac 12\frac 12\kappa], [\frac
12\kappa\frac 12]$ and $[\kappa\frac 12\frac 12],-\frac
12\leq\kappa\leq\frac 12$.  Thus, competing AFQ patterns create
fluctuations which should significantly decrease the AFQ-D transition
temperature, as compared to the mean-field estimate.  The wave vector
{\bf Q} is consistent with the AFQ patterns which have been found
experimentally by the Grenoble group \cite{effantin}.

At the two points of reciprocal space, namely, ${\bf 0}$ and {\bf Q},
we have ${\cal A}^{xx}={\cal A}^{zz}$.  Then the Fourier transform of
${\cal H}_{\rm orb}$ takes a planar form
$${\cal A}_{\bf k}(\tau_{-{\bf k},x}\tau_{{\bf k},x}+\tau_{-{\bf
k},z}\tau_{{\bf k},z})\,.$$ In other high-symmetry points, $[\frac
12\frac 120]$ and $[00\frac 12]$, ${\cal H}_{\rm orb}$ exhibits an
easy-axis form with non-equal values of ${\cal A}_{{\bf k}}^x$ and
${\cal A}_{{\bf k}}^z$.  It is also valid for a general point of
reciprocal space, but, in general, the off-diagonal component ${\cal
A}^{xz}$ is non-zero, and the easy-axis should be different from
either $x$-axis or $z$-axis.

It is worth noting, that searching for the ground state energy of the
{\it classical} vector field $\bbox\tau$ by using a Fourier
transformation of Hamiltonian (\ref{orb}) (also known as the
Luttinger-Tissa method) would be a standard procedure, if the
Hamiltonian were invariant under the homogeneous
$\bbox\tau$ rotations.  Nevertheless, although the
rotational symmetry of ${\cal H}_{\rm orb}$ at ${\bf k}=[\frac 12\frac
120]$ and ${\bf k}=[00\frac 12]$ is broken, the energy values listed
in Table I are rigorous.
\subsection{Magnetic ordering due to electric quadrupolar interactions}
In this section we consider a quantum effect, namely the zero
motion of the $\tau$ "spins" with respect to the AFQ background.  In
fact, when a decoupling procedure is applied to the ${\cal B}$ terms
in (\ref{qd1}), we obtain the effective spin-like Hamiltonian:
\begin{eqnarray}
\label{qd_m}
{\cal H}_{\rm m}=\frac 34
\sum_{{\bf r}\neq{\bf r}'}{\cal B}_{ij}({\bf r}\!-\!{\bf r}')
<\!\tau_y({\bf r})\tau_y({\bf r}')\!>\sigma_i({\bf r})
\sigma_j({\bf r}')
=\sum_{{\bf r}\neq{\bf r}'}
\widetilde{{\cal B}}_{ij}({\bf r}\!-\!{\bf r}')
\sigma_i({\bf r})\sigma_j({\bf r}')\,.
\end{eqnarray}
$\tau_y$ does not enter the Hamiltonian (\ref{orb}); this is
responsible for the formation of the orbital ordering.  This would be
a reason for neglecting all the contributions caused by $\tau_y$,
including interaction (\ref{qd_m}), were it not for quantum
fluctuations of $\bbox\tau$.  We put all the intermediate
formulae, which determine our choice of the quantization axis, the
spin-wave representation of $\tau$'s, etc., into Appendix C.

In the spin-wave approximation Hamiltonian (\ref{orb}) becomes
\begin{equation}
\label{sw2}
{\cal H}_{{\rm sw}}=\sum_{\bf q}\left(K_1({\bf q})b^{\dagger}_{\bf q}b_{\bf q}
+\frac 12 K_2({\bf q})(b_{-{\bf q}}b_{\bf q}+
b^{\dagger}_{\bf q}b^{\dagger}_{-{\bf q}})\right)\,,
\end{equation}
\vspace{-5mm}

\noindent
where
\begin{eqnarray}
\label{k2}
K_2({\bf q})=\frac 12 ({\cal A}^{zz}[\widetilde{\bf q}]\sin^2\phi
+{\cal A}^{xx}[\widetilde{\bf q}]\cos^2\phi-{\cal A}^{xz}[\widetilde{\bf q}]
\sin 2\phi),
\widetilde{\bf q}={\bf q}-{\bf Q},
\end{eqnarray}
and
\begin{eqnarray}
\label{k1}
K_1({\bf q})=K_2({\bf q})-{\cal A}_{\bf Q}\,.
\end{eqnarray}
For definition of $\phi$, see Appendix C.  The energy gain $E_{{\rm
sw}}^{(0)}$, which occurs due to the zero-motion of $\tau$'s is a
straightforward result of the Hamiltonian (\ref{sw2}) diagonalization:
\begin{eqnarray}
\label{sw3}
E_{{\rm sw}}^{(0)}=-\frac 12\sum_{\bf q}\left(K_1-\sqrt{K_1^2-K_2^2}\right)\,.
\end{eqnarray}
The correlation function 
\begin{eqnarray}
\label{tau_y}
\langle \tau_y(0)\tau_y({\bf r})\rangle=\frac 14 \sum_{\bf q}e^{i{\bf qr}}
\sqrt{\frac {K_1+K_2}{K_1-K_2}}
\end{eqnarray}
appears to be non-zero although it decays exponentially with distance $r$.

$E_{{\rm sw}}^{(0)}$ is a periodic function of $\phi$ with periodicity
$\pi/3$.  In fact, using equations (\ref{k2})-(\ref{sw3}) and
definitions of ${\cal A}$'s given in Appendix B, one can rigorously
prove that under the transformation $(q_x, q_y, q_z)\to (q_z, q_x,
q_y)$ ~$K_1({\bf q})$ and   
$K_2({\bf q})$ remain unchanged, if $\phi$
is simultaneously shifted by $\pi/3$.

Numerical calculations show that $\phi=0,\pi/3,2\pi/3$, etc., are
related to equivalent minimums of $E_{{\rm sw}}^{(0)}$.  Using one of
them, say, at $\phi=0$, we calculate the correlation functions
(\ref{tau_y}) numerically.  The sign of the first neighbor correlators
is negative: $\langle \tau_y(0)\tau_y({\bf a}_x)\rangle= \langle
\tau_y(0)\tau_y({\bf a}_y)\rangle\approx -0.0293$, $\langle
\tau_y(0)\tau_y({\bf a}_z)\rangle\approx -0.0058$.  This is a reminder
of the AFQ ordering.  Among the second neighbors only $\langle
\tau_y(0)\tau_y(\pm{\bf a}_x\pm{\bf a}_y)\rangle\approx -0.0082$ are
not negligible, all others are much smaller.  For the resulting
coupling constants of (\ref{qd_m}) see Appendix B.

This curious mechanism which, in principle, leads to the effective
magnetic interactions (see (\ref{qd_m})), could be a reason for
magnetic ordering at temperatures much smaller than $T_Q$, because,
{\it first}, the ${\cal B}$ coupling constants of Hamiltonian
(\ref{qdk}) are much less numerically than their ${\cal A}$
counterparts, and, {\it second}, the additional smallness comes from
the quantum fluctuations of $\tau_y$'s.  The low-temperature magnetism
in CeB$_6$ is unlikely to be described by such an unusual mechanism.
Such a mechanism would come into play only when all other magnetic
interactions (mainly via conductivity electrons) were very weak.
\section{AF-Quadrupolar -- Disorder Transition}
In this Section we consider CeB$_6$ near the AFQ-D transition.  For
this we employ, following Ref. \cite{uim1}, the spherical model which
is applicable to a system with well-pronounced soft modes.  The purpose
of this section is to determine
\begin{itemize}
\item
the shape of the AFQ-D transition line in the $T-H$ phase diagram;
\item
the singularity of the specific heat along this line.
\end{itemize}
\subsection{Spherical model and AFQ-D transition line}
From the behavior of the specific heat anomaly \cite{fujita,peysson}
(which is tiny in the weak magnetic field region) a strong short-range
AFQ order should exist above $T_Q(H)$.  A magnetic field suppresses
the fluctuations and makes $T_Q(H)$ higher.  In order to pick up these
features, we go beyond the mean-field approximation for Hamiltonian
${\cal H}_Z+{\cal H}_{\rm orb}$.  The first step in this direction
will be generalization of the spherical model for two spins,
$\bbox\sigma$ and $\bbox\tau$.  For taking into
account the quantum effects, we impose the constraints
$<\!{\bbox\sigma}^2({\bf r})\!>=\!3/4$ and
$<\!{\bbox\tau}^2({\bf r})\!>=\!1/2$.  The latter would be
equal to 3/4 were it not for redundancy of the $\tau_y$ variable.
Note, that, as shown in Ref. \cite{Hubbard}, the decoupling of
fluctuations in the spin-1/2 Heisenberg model leads to the spherical
model with the constraint 
$<\!{\bbox\sigma}^2({\bf
r})\!>=\!3/4$.  Now the partition function reads
\begin{eqnarray}
{\cal Z}=\!
\prod_{{\bf r}}\!{\displaystyle \int_{-\infty}^{\infty}}\!
d {\bbox\sigma}({\bf r})\!
{\displaystyle \int_{-\infty}^{\infty}}\!
d {\bbox\tau}({\bf r})
\exp \beta\{\lambda_{\sigma}
(3/4-{\bbox\sigma}^2({\bf r}))
+\lambda_{\tau}
(1/2-{\bbox\tau}^2({\bf r}))
-({\cal H}_{\rm orb}+{\cal H}_Z)\}
\label{zz}
\end{eqnarray}
where the spherical conditions 
$$
\partial {\cal F}/\partial \lambda_{\tau} =
\partial {\cal F}/\partial \lambda_{\sigma} = 0\quad ({\cal F}=-T\ln{\cal Z}),
$$
which control the constraints, should be satisfied by an appropriate choice 
of $\lambda_{\tau}$ and $\lambda_{\sigma}$. 
Gaussian integration over ${\bbox\sigma}(\bf r)$ and 
${\bbox\tau}(\bf r)$ in (\ref{zz}) is straightforward, that allows 
us to derive the free energy ${\cal F}$ of the spherical model.

For definiteness, we inspect the particular case of ${\bf H}\| [001]$:
This orientation is expected to favor the re-entrance of the AFQ--D
transition line at smaller $H$ as compared to other orientations.  The
singularities on the AFQ--D transition line, as well as its shape,
will be examined as temperature decreasing, {\it i.e.}, from the side
of the D-phase.

Performing the routine calculations, which are given in Appendix D,
we get (cf., (\ref{f0}),(\ref{f1s}),(\ref{f1t})):
\begin{eqnarray}
\label{f}
-{\cal F}=\frac 34\lambda_{\sigma}+\frac 12\lambda_{\tau}+
\left(\frac 78\right)^{\!2}
\frac{z(\lambda_{\tau}\!+\!{\cal A}_{\bf 0})}
{\lambda_{\tau}\!+\!{\cal A}_{\bf 0}\!-\!z}
+\frac 32 T\ln\frac{\pi T}{\lambda_{\sigma}}
\nonumber\\
+\frac T2\int\frac{d^3 k}{(2\pi)^3}
\ln\frac{(\pi T)^2}{(\lambda_{\tau}+{\cal A}^{zz}[{\bf k}]-z)
(\lambda_{\tau}+{\cal A}^{xx}[{\bf k}])-
({\cal A}^{xz}[{\bf k}])^2},
\end{eqnarray}
where
\begin{equation}
\label{z_h2}
z=\left(\frac 87\right)^{\!2}\frac {(\mu_BH)^2}{\lambda_{\sigma}}.
\end{equation}
Two equations,
\begin{equation}
\frac 12 -\left(\frac 78\right)^{\!2}\frac{z^2}
{(\lambda_{\tau}\!+\!{\cal A}_{\bf 0}\!-z)^2}
=\frac T2\int \frac{d^3k}{(2\pi)^3}
\frac{2\lambda_{\tau}+{\cal A}^{zz}[{\bf k}]+{\cal A}^{xx}[{\bf k}]-z}
{(\lambda_{\tau}+{\cal A}^{zz}[{\bf k}]-z)
(\lambda_{\tau}+{\cal A}^{xx}[{\bf k}])-
({\cal A}^{xz}[{\bf k}])^2}
\label{eq.1}
\end{equation}
and
\begin{eqnarray}
\frac 34\!-\!\left(\frac 78\right)^{\!2}\frac z{\lambda_{\sigma}}\!
-\!\left(\frac 78\right)^{\!2}\frac{z^2}
{\lambda_{\sigma}(\lambda_{\tau}\!+\!{\cal A}_{\bf 0}\!-z)}
\left(2\!+\!\frac{z}{\lambda_{\tau}\!+\!{\cal A}_{\bf 0}\!-z}\right)\!-\!
\frac{3T}{2\lambda_\sigma}
\nonumber\\
=\frac T2\frac z{\lambda_{\sigma}}\int \frac{d^3k}{(2\pi)^3}
\frac{\lambda_{\tau}\!+\!{\cal A}^{xx}[{\bf k}]}
{(\lambda_{\tau}\!+\!{\cal A}^{zz}[{\bf k}]\!-\!z)
(\lambda_{\tau}\!+\!{\cal A}^{xx}[{\bf k}])\!-\!
({\cal A}^{xz}[{\bf k}])^2},
\label{eq.2}
\end{eqnarray}
are valid for the D-phase and determine $\lambda_{\sigma}$ and
$\lambda_{\tau}$ as functions of $T$ and $H$, through which the
physical quantities can be then expressed.  The integrals in the
r.h.s. of Eqs.(\ref{eq.1})-(\ref{eq.2}) are not of that simple form as
the Watson integral, 
which appears in the spherical model treatment of
the 3D ferromagnet, but let us call them generalized Watson integrals.
The AFQ-D transition line corresponds to the values of
$\lambda_{\tau}$ at which the denominator of the generalized Watson
integrals turns zero at ${\bf k}={\bf Q}$.  Taken at the critical
value
\begin{eqnarray}
\label{cr_l}
\lambda_{\tau}^{(c)}(z)=|{\cal A}_{\bf Q}|+z\,, 
\end{eqnarray}
Eqs.(\ref{eq.1})-(\ref{eq.2}) determine $T_Q(H)$. 
This line is depicted in Fig. 5. 

At zero magnetic field, $\lambda_{\sigma}=2T_Q$ and $T_Q=(W_1(z=0))^{-1}$, 
as seen from Eqs.(\ref{eq.2}) and (\ref{eq.1}), respectively. 
We denote the 1$^{\rm st}$ generalized Watson integral (see
Eq.(\ref{eq.1})) taken along the transition line by $W_1(z)$. 
It can be written now as
\begin{eqnarray}
\label{i_1}
W_1(z)=\int\frac{d^3k}{(2\pi)^3}
\frac{2|{\cal A}_{\bf Q}|+{\cal A}^{zz}[{\bf k}]+{\cal A}^{xx}[{\bf k}]
+z}{(|{\cal A}_{\bf Q}|+
{\cal A}^{xx}[{\bf k}]+z)(|{\cal A}_{\bf Q}|+{\cal A}^{zz}[{\bf k}])-
({\cal A}^{xz}[{\bf k}])^2}.
\end{eqnarray}

In order to estimate the importance of fluctuations, we can compare
the spherical model and mean-field results for $T_Q$.  The mean-field
approach yields $T_Q^{MF}(H\!=\!0)=|{\cal A}_{\bf Q}|/2 = 5.37$, and
$T_Q(H)$ decreases monotonically with increasing $H$ \cite{uim1}.

An important property of $T_Q(H)$ should be mentioned in connection
with Fig. 5 (cf. Fig. 3), {\it i.e.}, $T_Q(H)$
grows {\it linearly} with $H$ at not very high magnetic fields.  This
feature is consistent with experimental findings.  
Mathematically,
this behavior follows from the properties of the generalized Watson
integrals: their expansions in a small parameter $z$ are
$c_0^{(i)}-c_1^{(i)}\sqrt z$, where all four values of $c$ are
positive.  Then, because in the leading order
$T_Q(H)-T_Q(0)\approx(W_1(0)-W_1(z))/(W_1(0))^2$, as it follows from
Eq.(\ref{eq.1}), the weak-field behavior of $T_Q$ becomes clearly
understood.  To complete this proof, let us represent
$(W_1(z)-W_1(0))$ as follows:
\begin{eqnarray}
\label{i_1_0}
W_1(0)-W_1(z)=z\int\frac{d^3k}{(2\pi)^3}\frac{
(|{\cal A}_{\bf Q}|+{\cal A}^{zz}[{\bf k}])^2+({\cal A}^{xz}[{\bf k}])^2}
{{\cal D}([{\bf k}];z){\cal D}([{\bf k}];z=0)},
\end{eqnarray}
which is positive.  We denote the denominator of integral (\ref{i_1})
by ${\cal D}([{\bf k}];z)$.  At $z=0$, integral in (\ref{i_1_0})
becomes singular at small $\delta{\bf k}={\bf k}-{\bf Q}$.  In fact,
the numerator behaves as $\delta{\bf k}^4$, whereas the denominator
$\propto \delta{\bf k}^8$.  The integral would diverge as $|\delta{\bf
k}|^{-1}$ were it not for the cut-off at $|\delta{\bf k}|^2\sim z$.
Thus we arrive at $W_1(0)-W_1(z)\propto \sqrt{z}$.

It is worth noting, that the spherical model is a reasonable tool for
picking up strong fluctuations. However if fluctuations are effectively
suppressed, as is the case in high magnetic fields, the spherical model
performs less satisfactorily.
\subsection{Specific heat}
Here we confine our consideration to the vicinity of the AFQ-D
transition line and magnetic fields weak as compared to $T_Q$
($\mu_BH\ll T_Q$).  The equations which determine thermodynamic
behavior are (\ref{eq.1})-(\ref{eq.2}), where we can neglect all high
order terms in $H$ ($H^n$ with $n\geq 2$).  Using the spherical
conditions ($\partial {\cal F}/\partial \lambda_\sigma= \partial {\cal
F}/\partial \lambda_\tau=0$), one can obtain for the entropy
($S=-\partial {\cal F}/\partial T$):
\begin{eqnarray}
\label{ent}
S=\frac 32\ln\frac{\pi T}{\lambda_{\sigma}}\!+\!\frac 52\!+\!
\frac 12\int\frac{d^3k}{(2\pi)^3}
\ln\frac{(\pi T)^2}{(\lambda_{\tau}\!+\!{\cal A}^{zz}[{\bf k}]\!
-\!z)(\lambda_{\tau}\!+\!{\cal A}^{xx}[{\bf k}])\!-
\!({\cal A}_{xz}[{\bf k}])^2}.
\end{eqnarray}
The specific heat at constant field, $C_H=T\,(\partial S/\partial T)_H$, 
is now determined as follows
\begin{eqnarray}
C_H=\frac 52-\frac T2\frac{d\lambda_{\tau}}{dT}\int
\frac{d^3k}{(2\pi)^3}\frac{2\lambda_{\tau}+
{\cal A}^{zz}[{\bf k}]+{\cal A}^{xx}[{\bf k}]-z}{
(\lambda_{\tau}+{\cal A}^{zz}[{\bf k}]
-z)(\lambda_{\tau}+{\cal A}^{xx}[{\bf k}])-({\cal A}_{xz}[{\bf k}])^2}
\nonumber\\
\label{sh}
-\frac 32\frac T{\lambda_{\sigma}}\frac{d\lambda_{\sigma}}{dT}
-\frac {Tz}{2\lambda_{\sigma}}\frac{d\lambda_{\sigma}}{dT}
\int\frac{d^3k}{(2\pi)^3}\frac{\lambda_{\tau}+
{\cal A}^{xx}[{\bf k}]}{
(\lambda_{\tau}+{\cal A}^{zz}[{\bf k}]
-z)(\lambda_{\tau}+{\cal A}^{xx}[{\bf k}])-({\cal A}_{xz}[{\bf k}])^2}\,.
\end{eqnarray}
With using Eqs.(\ref{eq.1})-(\ref{eq.2}) we can transform (\ref{sh})
into the simple form:
\begin{eqnarray}
\label{sh1}
C_H=\frac 52-\frac 12\frac {d\lambda_{\tau}}{dT}-
\frac 34\frac {d\lambda_{\sigma}}{dT}\,.
\end{eqnarray}
Let us consider a fixed magnetic field and $T\!>\!T_Q(H)$, {\it i.e.},
$T=T_Q(H)+\delta T$.  At $\delta T\ll T_Q$, $\lambda_{\tau}=|{\cal
A}_Q|+z+\delta \lambda_\tau$ is supposed to be slightly different from
$\lambda_{\tau}^{(c)}(z)$ (see (\ref{cr_l})).  In order to find out
$\delta \lambda_\tau (\delta T)$ we return to Eq.(\ref{eq.1}) which
takes the following form at $z\ll T_Q$:
\begin{eqnarray}
\label{sol1}
\frac 12=\frac 12(T_Q(H)+\delta T)(W_1(z)-\sqrt{\delta \lambda_\tau}
w(z/\delta \lambda_\tau;\delta \lambda_\tau)),
\end{eqnarray}
where $w$ is positive:
\begin{eqnarray}
w(z/\delta \lambda_\tau;\delta \lambda_\tau)
=\frac {W_1(z)}{\sqrt{\delta \lambda_\tau}}\nonumber\\
-\frac 1{\sqrt{\delta \lambda_\tau}}\!\int\!\frac{d^3k}{(2\pi)^3}
\frac{2|{\cal A}_{\bf Q}|+{\cal A}^{zz}[{\bf k}]+
{\cal A}^{xx}[{\bf k}]+2\delta \lambda_\tau+z}{(|{\cal A}_{\bf Q}|\!+\!
{\cal A}^{xx}[{\bf k}]\!+\!\delta \lambda_\tau\!+\!z)
(|{\cal A}_{\bf Q}|\!+\!{\cal A}^{zz}[{\bf k}]\!+\!\delta\lambda_\tau)\!-\!
{\cal A}^{xz}[{\bf k}]^2}
\label{w_f}
\end{eqnarray}
In Fig. 6 $w$ is depicted as function of $\delta \lambda_\tau$
for three values of the ratio $z/\delta \lambda_\tau$.  It can be
remarkably well approximated to the form $w_0(\delta
\lambda_\tau)\cdot (1+\alpha(\delta \lambda_\tau)\sqrt{z/\delta
\lambda_\tau})^{-1}$, where $w_0(x)=w(0;x)$ (see the upper curve in
Fig. 6) and $\alpha(x)$ are both weakly dependent functions of
the argument. For example, $\alpha(0)\approx 0.7$ and
$\alpha(1)\approx 1.15$. Note, that $\alpha$ and $wT_Q^{3/2}$ are
dimensionless quantities ($w_0T_Q^{3/2}\sim 1$).  Thus, in the leading
order, when $\delta \lambda_\tau\ll z$, which is equivalent to $\delta
T\ll \mu_B H$, we obtain:
$$
\delta \lambda_\tau\approx\frac{\alpha(0)}{w_0(0)}
\frac{W_1(z)}{T_Q}\,\delta T\sqrt{z}
$$
and, because $W_1(z)\approx T_Q^{-1}$,
\begin{eqnarray}
\label{dt}
\frac{d\lambda_{\tau}}{dT}\approx\frac{\alpha(0)}{w_0(0)}
\frac{1}{T_Q^2}\sqrt{z}.
\end{eqnarray}
\noindent
In accordance with Eq.(\ref{eq.2}) 
\begin{eqnarray}
\label{ls}
\lambda_{\sigma}=2T+O((\mu_BH)^2/T_Q). 
\end{eqnarray}
Therefore, in the leading order in $H$ one can obtain from (\ref{sh1}),
(\ref{dt}) and (\ref{ls}):
\begin{eqnarray}
\label{sh+}
\lim_{\delta T\rightarrow +0}C_H=1-\frac 47\frac{\alpha(0)}{w_0(0)}\frac 1{T_Q^2}
\frac{\mu_BH}{\sqrt{2T_Q}}\,.
\end{eqnarray}
Below $T_Q(H)$ $\lambda_{\tau}$ does not depend anymore on $\delta T$
and remains equal to $|{\cal A}_Q|+z$. 
Thus, we get
\begin{eqnarray}
\label{sh-}
\lim_{\delta T\rightarrow -0}C_H=1\,,
\end{eqnarray}
which, together with (\ref{sh+}), determines the specific heat jump, 
increasing linearly with $H$.

At $T_Q\gg \delta T\gg \mu_B H$, {\it i.e.}, beyond a narrow vicinity of the 
transition line, we arrive at $C_H$, decreasing linearly with $\delta T$:
\begin{eqnarray}
C_H=1-T_Q^{-1}w_0^{-1}\delta T
\nonumber
\end{eqnarray}
\section{Well-isolated $\Gamma_8$-levels: a single $f$-hole configuration}
\subsection*{ Zeeman and quadrupolar interactions}
Now we deal with the following quantum numbers: ${\cal J}\!=\!7/2$,
$S\!=\!1/2$ and $L\!=\!3$.  In the crystal field of cubic symmetry the
eightfold multiplet splits into the $\Gamma_8$ quartet and two
doublets, $\Gamma_6$ and $\Gamma_7$.  If the crystal field Hamiltonian
allows the quartet to be realized as a well-isolated ground state
level, then we confine our consideration to the following set of
wave-functions:
\begin{eqnarray}
\label{jhole}
\psi_{1,\pm}=\pm\sqrt{\frac{7}{12}}|\mp 7/2\rangle\mp
\sqrt{\frac{5}{12}}|\pm 1/2\rangle,\quad
\psi_{2,\pm}=\mp\frac 12|\pm 5/2\rangle\mp\frac{\sqrt{3}}2|\mp 3/2\rangle.
\end{eqnarray}   
Listed below in units of $\mu_B$ are the non-zero matrix elements of $M_z$, 
$M_+$ and $M_-$:
\begin{eqnarray}
\label{j8}
\begin{array}{l}
<\!\psi_{1,+}|M_z|\psi_{1,+}\!>
=\!-\!<\!\psi_{1,-}|M_z|\psi_{1,-}\!>
=\!-\frac {44}{21};\\
<\!\psi_{2,+}|M_z|\psi_{2,+}\!>
=\!-\!<\!\psi_{2,-}|M_z|\psi_{2,-}\!>
=\!-\frac {4}{7};\\
<\!\psi_{2,-}|M_+|\psi_{1,+}\!>=
<\!\psi_{1,+}|M_-|\psi_{2,-}\!>=\!-\frac{16}{7\sqrt{3}};
\\
<\!\psi_{2,+}|M_-|\psi_{1,-}\!>=
<\!\psi_{1,-}|M_+|\psi_{2,+}\!>=\!-
\frac{16}{7\sqrt{3}};\\
<\!\psi_{1,-}|M_-|\psi_{1,+}\!>=
<\!\psi_{1,+}|M_+|\psi_{1,-}\!>=
\!-\frac {40}{21};\\
<\!\psi_{2,-}|M_-|\psi_{2,+}\!>=
<\!\psi_{2,+}|M_+|\psi_{2,-}\!>=
\!-\frac {24}7.
\end{array}
\end{eqnarray}
All these matrix elements are in accordance with the operator expression:
\begin{equation}
\label{mom_h}
M_i=-\frac 83\mu_B\sigma_i(1+\frac 87T_i)\,,\quad i=x,y,z.
\end{equation}
Note, that the only difference between Eqs.(\ref{mom_h}) and (\ref{mom})
is the coefficient (-- 8/3 instead of 2).

Now let us determine the matrix $\|Q\|$. 
For the unit, we take
$$
Q_0=<\!\psi_{1,\sigma}|Q_{zz}|\psi_{1,\sigma}\!>.
$$
The diagonal components of the quadrupolar moment have the same
non-zero matrix elements as in Eq.(\ref{q_ind}), whereas the matrix
elements of the off-diagonal components are six times larger than the
corresponding elements of (\ref{q_dep}).  Thus, the form of $\|Q\|$ is
the same as in (\ref{q_matr}), but the off-diagonal "vectors", which
transform in accordance with $\Gamma_5$, are defined now as
${\bbox\mu}=3\sqrt{3}\tau_y{\bbox\sigma}$.  This
circumstance makes the problem of quadrupolar ordering less
transparent as in the single $f$-electron case: The part of ${\cal
H}_{\rm qd}$ (see Eq.(\ref{qd1})), which reflects the $\mu-\mu$
interactions, becomes as important as the $\tau-\tau$ part.

There are a few compounds of cubic symmetry, based on the single
$f$-hole ions, which can be the candidates to realize the $\Gamma_8$
quadrupolar ordering.  Among them we would mention YbB$_{12}$
\cite{sugi} and TmTe \cite{matsum}.  There is another important
difference between these face-cubic centered compounds and CeB$_6$
(recall, that Ce ions are arranged in a simple cubic lattice).
Namely, fcc compounds do not exhibit such pronounced soft modes as in
the simple cubic case.  It is not our purpose to give a detailed
analysis of the fcc situation in the framework of the realistic
quadrupolar interaction (\ref{v}).  We mention, however, that the
ground state of the analog of ${\cal H}_{\rm orb}$ (see (\ref{orb}))
is of the AFQ-type, which is related to the high-symmetry points
(${\bf Q}_x, {\bf Q}_y, {\bf Q}_z$) of the Brillouin zone boundaries
(see Fig. 7): $E({\bf k}={\bf Q}_i)\approx -8.85$ at them.
The order parameter at, say ${\bf k}={\bf Q}_z$, corresponds to the
Ising-like symmetry with $\langle\tau_z\rangle=0$ and
$\langle\tau_x\rangle$, altering the sign from layer to layer.  The
direction of low-lying excitations coincides with [001], but the mode
is not a soft one: $E({\bf k}={\bf 0})-E({\bf k}={\bf Q})\approx
3.99$.
\section{Discussion and Conclusions}
The problem of quadrupolar ordering in CeB$_6$ seems to be
well-defined provided we confine our attention to the Zeeman energy
and direct quadrupolar interactions.  The unusual form of these Zeeman
and quadrupolar terms owes to the well-isolated $\Gamma_8$ quartet.
Instead of dealing with Stevens operators, it is more convenient to
introduce the spin-like, $\bbox\sigma$, and orbital-like,
$\bbox\tau$, operators.  However, in the low-temperature
and weak-field region CeB$_6$ undergoes the magnetic phase transition
(see Fig. 2), which results in the appearance of
complicated magnetic structures.  For their explanation it is
insufficient to restrict ourselves to the above-mentioned interactions
only, but indirect interactions via conductivity electrons start
playing an important role.  Magnetic domains of different orientations
have been identified in neutron diffraction experiments \cite{thes}
for magnetic fields applied along [111], [110] and [001].  However the
interpretation of neutron experiments \cite{effantin,erk,thes} occurs
to be contradictory to recent $\mu$SR measurements \cite{feyer}.  As
for the AFQ ordered state, the neutron NMR results are still in
disagreement: The triple-${\bf k}$ structure has been proposed by
Takigawa {\it et al.} \cite{takigawa}, whereas in all the neutron
experiments the {\bf Q} modulation has been reported (${\bf
Q}\!=\![\frac 12\frac 12\frac 12]$).  These two interpretations
are mutually exclusive.  Unfortunately, Ref. \cite{takigawa} is a short
paper with not many details, we mention a few of them to show a
significant difference of the triple-${\bf k}$ ({\bf q}$_1\!=\![\frac
1200]$, {\bf q}$_2\!=\![0\frac 120]$, {\bf q}$_3\!=\![00\frac 12]$)
and ${\bf Q}$ modulated structures.
For the modulated component of magnetization ${\bf m}({\bf r})$
the following equation has been proposed in \cite{takigawa}:
\begin{eqnarray}
{\bf m}({\bf r})=(-1)^{\ell}{\bf m}_1+(-1)^m{\bf m}_2+(-1)^n{\bf m}_3,
\label{tak1}
\end{eqnarray}
where {\bf r}=($\ell,m,n$), and the polarization vectors depend on the 
magnetic field direction (${\bf H}\|(\cos\theta_1,\cos\theta_2,\cos\theta_3)$)
through the equations:
\begin{eqnarray}
\begin{array}{l}
{\bf m}_1=m_1(\theta_1)(\cos \theta_1, -\cos \theta_2, -\cos \theta_3)
\\
{\bf m}_2=m_2(\theta_2)(-\cos \theta_1, \cos \theta_2, -\cos \theta_3)
\\
{\bf m}_3=m_3(\theta_3)(-\cos \theta_1, -\cos \theta_2, \cos \theta_3)
\end{array}
\label{tak2}
\end{eqnarray}
An important property of \{$m_i$\} is that $m_i(\pi/2)=0$.   
The form of Eqs.(\ref{tak1})-(\ref{tak2}) is not transparent, it is easier to 
illustrate them with a couple of examples.

In the first example {\bf H}$\|[001]$, which leads to
$\theta_1=\theta_2=\pi/2$, hence, $m_1=m_2=0$.  This corresponds to a
degeneracy of the triple-${\bf k}$ structure, whose realization now is
a single-{\bf k} structure.  According to (\ref{tak2}) the modulated
component of magnetization is arranged as shown in Fig. 8a.
Our theoretical description results in the arrangement shown in Fig.
8b.

In the second example {\bf H}$\|[110]$, and $\theta_3=\pi/2,\;
\theta_1=\theta_2=\pi/4$ lead to $m_3=0,\; m_1=m_2$.  This degeneracy
corresponds to a double-{\bf k} structure.  Figs. 9a and
9b are the NMR and theoretical interpretations,
respectively.  Note, that the Fig. 9a pattern reproduces
itself under any translation along [001], while {\bf m}{\footnotesize
s} of Fig. 9b alter the sign under the translation by the
elementary lattice constant.

The preliminary $\mu$SR results \cite{feyer} contradict both
neutron and NMR, measurements.

A critical experiment seems to be not difficult to achieve.
It is connected with the X-Ray structural measuments at {\it zero}
magnetic field.  The translational electric symmetry of the Ce-lattice
is broken below $T_Q$: although all the electric charges of Ce ions
are equal to each other, the modulated component of the quadrupolar
moment should contribute to the Bragg reflections at the transferred
wave vectors of the $[\frac 12\frac 12\frac 12]$-type.  This would be
a weak effect, caused by only one electron of the total number $Z$.
However, instead of an usual X-Ray technique it would be possible to
use synchrotron facilities for finding the {\bf Q} or triple-{\bf k}
modulated structure (or something different from these two).  Note,
that the zero field experiment is more instructive because it allows
to avoid the secondary effects in non-zero fields, {\it i.e.}
formations of magnetically modulated structures.  To complete this
X-Ray discussion we give the on-site form-factor {\it operator} which
is calculated over the set $\{\psi_{\ell,\sigma}\}$ and related to the
$f$ electron {\it only}:
\begin{eqnarray}
f({\bf q})=<\!j_0\!>+\frac 12\!<\!j_4\!>\!P_4(z_{\bf q})
+\frac 17\tau_z\,({\bf r})(16\!<\!j_2\!>\!P_2(\cos\theta_{\bf q})-
5\!<\!j_4\!>\!P_4(z_{\bf q}))\nonumber\\
+\frac 1{7\sqrt{3}}\tau_x\,({\bf r})(8\!<\!j_2\!>\!P_2^2(z_{\bf q})+\!<\!j_4\!>\!P_4^2(z_{\bf q}))
\cos 2\phi_{\bf q},\quad z_{\bf q}=\cos\theta_{\bf q}.\nonumber
\end{eqnarray}
$\theta_{\bf q}$ and $\phi_{\bf q}$ denote the spherical coordinates
of {\bf q} relative to the $z$-axis.  $P_l^m(z)$ are associated
Legendre polynomials.  Detailed calculations in connection with a
concrete X-Ray (synchrotron) experiment will be published elsewhere.

The experimental technique which is associated with the so-called
3$^{\rm rd}$ order paramagnetic susceptibility \cite{mor2} can be also
used for probing the quadrupolar ordering in CeB$_6$.  However, it
cannot yield an information about the microscopic arrangement of
quadrupolar moments.  This method is based on the extraction of the
$H^3$-terms from $M(H)$, when $H$ is small.  For our problem, the
average magnetization can be written as:
\begin{eqnarray}
\label{3rd_g}
\overline{M}_\alpha=-\frac 1{6T^3}\frac 1N\sum_{{\bf r}_1\dots {\bf r}_4}\{
\langle M_\alpha({\bf r}_1)M_\beta({\bf r}_2)M_\mu({\bf r}_3)M_\nu({\bf r}_4)
\rangle
\nonumber\\
-3\langle M_\alpha({\bf r}_1)M_\beta({\bf r}_2)\rangle
\langle M_\mu({\bf r}_1)M_\nu({\bf r}_2)\rangle\}H_\beta H_\mu H_\nu
\end{eqnarray}
For further transformations in (\ref{3rd_g}), we imply that, {\it
first}, all the magnetic fluctuations are much smaller as compared to
the quadrupolar fluctuations in the AFQ phase (at least near $T_Q$),
{\it second}, the diagonal components of the quadrupolar moment are
responsible for ordering below $T_Q$.  Then, using the Wigner-Eckhart
theorem we can decouple Eq.(\ref{3rd_g}) according to the following
scheme:
$$
M_\alpha({\bf r})M_\beta({\bf r}')\to 
\left(\frac 76\right)^{\!2}\mu_B^2{\cal J}_\alpha({\bf r})
{\cal J}_\beta({\bf r})\delta_{{\bf r},{\bf r}'}\to 
\left(\frac 76\right)^{\!2}\mu_B^2 Q_0^{-1}(T_{\alpha}({\bf r})+{\cal J}
({\cal J}\!+\!1)/3)\delta_{\alpha\beta}\delta_{{\bf r},{\bf r}'}.
$$
and arrive at the following equation $\overline{M}_\alpha=
\overline{M}_\alpha^{(0)}+\overline{M}_\alpha^{(1)}$, where
\begin{eqnarray}
\label{3rd_1}
\overline{M}_\alpha^{(1)}=-\frac {\mu_B^4}{2T^3}\left(\frac 76\right)^{\!4}
\frac 1N
\sum_{{\bf r}'\neq {\bf r}}\frac 1{Q_0^2}\sum_{\mu}
(\langle T_\alpha({\bf r})T_\mu({\bf r}')\rangle -
\langle T_\alpha({\bf r})\rangle \langle T_\mu({\bf r}')\rangle)
H_\alpha H_\mu^2,
\end{eqnarray}
\begin{eqnarray}
\label{3rd_0}
\overline{M}_\alpha^{(0)}=-\frac {\mu_B^4}{6T^3}\left(\frac 76\right)^{\!4}
\frac 1N\sum_{\bf r}\sum_{\beta,\mu,\nu} H_\beta H_\mu H_\nu
(\langle {\cal J}_\alpha({\bf r}){\cal J}_\beta({\bf r})
{\cal J}_\mu({\bf r}){\cal J}_\nu({\bf r})\rangle\nonumber\\
-\langle{\cal J}_\alpha({\bf r}){\cal J}_\beta({\bf r})\rangle
\langle{\cal J}_\mu({\bf r}){\cal J}_\nu({\bf r})\rangle\!-\!
\langle{\cal J}_\alpha({\bf r}){\cal J}_\mu({\bf r})\rangle
\langle{\cal J}_\beta({\bf r}){\cal J}_\nu({\bf r})\rangle\!-\!
\langle{\cal J}_\alpha({\bf r}){\cal J}_\nu({\bf r})\rangle
\langle{\cal J}_\beta({\bf r}){\cal J}_\mu({\bf r})\rangle)
\end{eqnarray}
Eq.(\ref{3rd_1}) includes the irreducible correlators of $\tau_x$ and
$\tau_z$. The on-site irreducible correlator enters Eq.(\ref{3rd_0}).
The average of four ${\cal J}$'s can be reduced to a linear-in-$\tau$
expectation value which disappears upon summation over {\bf r}.  The
$<\!{\cal J}{\cal J}\!>^2$ terms of (\ref{3rd_0}) result in the
contribution $\propto \langle \tau\rangle^2$, which should produce a
kink in a dependence $\chi^{(3)}$ versus $T$ at $T_Q$.

The AFQ-D transition line was recently measured in fields up to 18
Tesla \cite{lacerda}.  In spite of its tendency to re-enter, this
curve still displays the monotonic $T_Q(H)$ behavior. An optimistic
theoretical prediction is $\sim$ 25--30 Tesla for the field at which
the re-entrance could start and $\sim$ 80 Tesla for the
zero-temperature critical field.  The current experimental facilities
are enough to examine the field region around 25--30 Tesla.

Although many experimental results can be explained by the present
theory (see also \cite{uim1,uim2}), still there remain a few puzzling
facts.  Among them we would mention the AFQ-D transition line whose
experimental shape is practically independent on the field
orientation, [001], [110] or [111].  Probably, such behavior could be
ascribed to an unusual anisotropy of the Zeeman energy.  In fact, if
we consider an isolated $\Gamma_8$ ion, its ground state energy
depends on the magnetic field direction as follows:
\begin{eqnarray}
\label{grst}
E_{\rm gs}=-\frac {\mu_BH}7\sqrt{65+4\sqrt{270(n_x^4+n_y^4+n_z^4)-74}},
\end{eqnarray}
that is -11, -9.81 and -9 (in units $\mu_BH/7$) for orientations
[001], [110] and [111], respectively.  For magnetic field of a general
orientation, the vector of magnetization in such a paramagnetic state
does not follow the same direction.  In this connection, the
experiments with diluted compounds La$_{1-x}$Ce$_x$B$_6$ could provide
an important information, if the crystal field still favors the
$\Gamma_8$ ground state.

From the theoretical point of view it should be interesting to
understand the symmetry and a microscopic origin of the magnetic
interactions which govern the properties of the system at low
temperatures.
\subsection*{Acknowledgement} 
I take the opportunity to thank A. Lacerda and M. Torikachvili for
sending me the experimental data prior to their publication.  When
making this work I had fruitful and enlightening discussions with
P. Burlet, H. Capellmann, Yu. Chernenkov, J. Flouquet, T. Kasuya,
V. Mineev, P. Morin, E. M\"uller-Hartmann, V. Plakhty, L.-P. Regnault,
J. Schweizer, C. Vettier.  It is my pleasure to thank M. Burgess for 
linguistic comments. This work has been done during my stay at
CEN/CNRS in Grenoble, in accordance with the program of the Ecole Normale
Superieure -- Landau Institute cooperation.
\appendix 
\subsection*{Appendix A}
In order to construct the operator expressions for ${\bf M}$, as well
as for $Q_{ij}$, we employ Table II.  It shows the operator connection
between all four possible states. 
Using matrix elements (\ref{j1}) in
combination with Table II one can obtain for $M_x=(M_++M_-)/2$, for
example:
\begin{eqnarray}
\label{mx}
M_x=\frac{\mu_B}{14}(4\sqrt{3}
(\tau_-\sigma_-+\tau_+\sigma_++\tau_+\sigma_-+\tau_-\sigma_+)+
10(1/2+\tau_z)(\sigma_++\sigma_-)\nonumber\\
+18(1/2-\tau_z)(\sigma_++\sigma_-))
=2\mu_B\sigma_x(1+\frac 47(\sqrt{3}\tau_x-\tau_z))
\end{eqnarray}
Another example is for $Q_{xy}$ (see Eqs.(\ref{q_dep})):
\begin{eqnarray}
\label{qxy}
Q_{xy}= Q_0\left(i\frac{\sqrt{3}}8(\tau_-
(1/2+\sigma_z)+\tau_+(1/2-\sigma_z)\right.
\nonumber\\
-i\left.\frac{\sqrt{3}}8(\tau_+(1/2+\sigma_z)+\tau_-(1/2-\sigma_z))\right)=
\frac {\sqrt{3}}2\sigma_z\tau_y=\mu_z
\end{eqnarray}
\subsection*{Appendix B}
The original parameters ${\cal A}_{ij,mn}({\bf r})$ give rise to 
${\cal A}_{\alpha\beta}({\bf r})$ and ${\cal B}_{ij}({\bf r})$ whose angular 
dependence is derived below
in accordance with Eqs.(\ref{q_matr},\ref{qd},\ref{qd1}):
\begin{eqnarray}
{\cal A}_{zz}\!=\!{\cal A}_{xx,xx}\!+\!{\cal A}_{yy,yy}\!+\!4{\cal A}_{zz,zz}\!+\!2{\cal A}_{xx,yy}\!-\!4{\cal A}_{xx,zz}\!-\!4{\cal A}_{yy,zz}
\!=\!\frac {35}{24}(1\!-\!3n_z^2)^2\!+\!\frac 56 (1\!-\!3n_z^2)\!-\!\frac 76, 
\nonumber\\
{\cal A}_{xx}\!=\!3{\cal A}_{xx,xx}\!+\!3{\cal A}_{yy,yy}\!-\!6{\cal A}_{xx,yy}
\!=\!\frac {35}{8}(n_x^2\!-\!n_y^2)^2\!-\!\frac 56 (1\!-\!3n_z^2)\!-\!
\frac 76, \nonumber\\
{\cal A}_{xz}\!=\!\sqrt{3}\,(-{\cal A}_{xx,xx}\!+\!{\cal A}_{yy,yy}\!+\!2{\cal A}_{xx,zz}\!-\!2{\cal A}_{yy,zz})
\!=\!\sqrt{3}\,(n_x^2\!-\!n_y^2)\left(-\frac {35}{24}(1\!-\!3n_z^2)\!+\!\frac 56\right).
\nonumber
\end{eqnarray}
\begin{eqnarray}
{\cal B}_{xx}\!=\!4{\cal A}_{yz,yz}\!=\!\frac {35}6n_y^2n_z^2\!+\!\frac 56
n_x^2\!-\!\frac 23,\quad {\cal B}_{yy}\!=\!4{\cal A}_{zx,zx}\!=\!\frac {35}6n_z^2n_x^2\!+\!\frac 56n_y^2\!-\!\frac 23,\nonumber\\
{\cal B}_{zz}\!=\!4{\cal A}_{xy,xy}\!=\!\frac {35}6n_x^2n_y^2\!+\!\frac 56
n_z^2\!-\!\frac 23,\quad 
 {\cal B}_{xy}\!=\!4{\cal A}_{yz,xz}\!=\!\frac 56n_xn_y(7n_z^2\!-\!1),
\nonumber\\
{\cal B}_{yz}\!=\!4{\cal A}_{zx,xy}\!=\!\frac 56n_yn_z(7n_x^2\!-\!1),\quad 
{\cal B}_{zx}\!=\!4{\cal A}_{xy,yz}\!=\!\frac 56n_zn_x(7n_y^2\!-\!1).
\nonumber
\end{eqnarray}
To shorten the ${\cal A}$ and ${\cal B}$ expressions we skip their 
$r$-dependence, {\it i.e.} factor $r^{-5}$. 

Listed below are a few coupling constants of the effective $\sigma-\sigma$
Hamiltonian (\ref{qd_m}):
\begin{eqnarray}
\widetilde{\cal B}_{xx}({\bf a}_x)=0.0049, 
\quad\widetilde{\cal B}_{yy}({\bf a}_x)=
\widetilde{\cal B}_{zz}({\bf a}_x)=-0.0195\nonumber\\
\widetilde{\cal B}_{yy}({\bf a}_y)=0.0049, 
\quad\widetilde{\cal B}_{xx}({\bf a}_y)=
\widetilde{\cal B}_{zz}({\bf a}_y)=-0.0195\nonumber\\
\widetilde{\cal B}_{zz}({\bf a}_z)=-0.0010, 
\quad\widetilde{\cal B}_{xx}({\bf a}_z)=
\widetilde{\cal B}_{yy}({\bf a}_y)=0.0039\nonumber\\
\widetilde{\cal B}_{zz}({\bf a}_x\pm{\bf a}_y)=-0.0011,\quad
\widetilde{\cal B}_{xy}({\bf a}_x\pm{\bf a}_y)=\pm 0.0006\nonumber
\end{eqnarray}
\subsection*{Appendix C}
Because of the rotational symmetry of the {\bf Q}-modulated state,
we choose the quantization axis, ${\bf n}_0$, as lying in the 
($\tau_x$, $\tau_z$) plane. 
One of the two perpendicular directions, ${\bf n}_1$, is taken also in the 
($\tau_x$, $\tau_z$) plane. 
Let the third direction coincide with $\tau_y$:
\begin{eqnarray}
\begin{array}{lccccl}
{\bf n}_0=(\!\!\! & \sin\phi, & 0, &\!\! &\!\cos\phi &\!\!\!)\\  
{\bf n}_1=(\!\!\! & \cos\phi, & 0, &\!\!-&\!\sin\phi &\!\!\!)\\
{\bf n}_2=(\!\!\! & 0, & 1, &\!\! &\! 0 &\!\!\!)   
\end{array}
\nonumber
\end{eqnarray}
In the spin-wave approximation
\begin{eqnarray}
\begin{array}{l}
{\bbox\tau}\cdot{\bf n}_0=e^{i{\bf Qr}}\left(\frac 12 - b^{\dagger}b\right)\\
{\bbox\tau}\cdot{\bf n}_1=\frac 12 e^{i{\bf Qr}}(b+b^{\dagger})\\
{\bbox\tau}\cdot{\bf n}_2=\frac 1{2i}(b-b^{\dagger}) 
\end{array}
\nonumber
\end{eqnarray}
Hence, to the same order we obtain
\begin{eqnarray}
\label{tau_op}
\begin{array}{l}
\tau_x=e^{i{\bf Qr}}\left[\left(\frac 12-n\right)\sin\phi+
\frac12(b+b^{\dagger})\cos\phi\right]\\
\tau_z=e^{i{\bf Qr}}\left[\left(\frac 12-n\right)\cos\phi-
\frac 12(b+b^{\dagger})\sin\phi\right]\\
\tau_y=\frac 1{2i}(b-b^{\dagger})
\end{array}
\end{eqnarray}
Now operators (\ref{tau_op}) can be used for extracting  
the "classical" and "spin-wave" parts of ${\cal H}_{\rm orb}$:
\begin{eqnarray}
\label{sw1}
\sum_{{\bf r}\neq {\bf r}'}({\cal A}_{zz}({\bf r}-{\bf r}')
\tau_z({\bf r})
\tau_z({\bf r}')+{\cal A}_{xx}({\bf r}-{\bf r}')
\tau_x({\bf r})\tau_x({\bf r})+
{\cal A}_{xz}({\bf r}-{\bf r}')(\tau_x({\bf r})\tau_z({\bf r}')
+\tau_z({\bf r})\tau_x({\bf r}')))\nonumber\\
=\sum_{\bf q}(\frac 14-b^{\dagger}_{\bf q}b_{\bf q})
\sum_{\bf r}e^{i{\bf Qr}}({\cal A}_{zz}({\bf r})\cos^2\phi+\!{\cal A}_{xx}
({\bf r})\sin^2\phi+\!{\cal A}_{xz}({\bf r})\sin 2\phi)
\nonumber\\
+\frac 14\sum_{\bf q}\sum_{\bf r}e^{i{\bf Qr}}e^{i{\bf qr}}({\cal A}_{zz}
({\bf r})\sin^2\phi+ 
{\cal A}_{xx}({\bf r})\cos^2\phi-{\cal A}_{xz}({\bf r})\sin 2\phi)
(b_{-{\bf q}}+b^{\dagger}_{\bf q})(b_{\bf q}+b^{\dagger}_{-{\bf q}})\,.
\end{eqnarray}
The ``classical'', i.e., operator-independent, part of the r.h.s of 
Eq.(\ref{sw1}) is invariant under the rotation of ${\bf n}_0$ in the 
($\tau_z$, $\tau_x$) plane ($\phi$-rotation). In fact, this part
appears to be $\phi$-independent: 
$N {\cal A}_{\bf Q}/4.$
\subsection*{Appendix D}
Let us first divide both fields, $\bbox\sigma$ and 
$\bbox\tau$, into uniform and modulated parts:
$$
{\bbox\sigma}({\bf r})={\bbox\sigma}^{(0)}+
\widetilde{\bbox\sigma}({\bf r}),\quad 
{\bbox\tau}({\bf r})=
{\bbox\tau}^{(0)}+\widetilde{\bbox\tau}({\bf r}).
$$
Then, the coefficients of the linear in $\widetilde{\bbox\sigma}$ 
and $\widetilde{\bbox\tau}$ terms in the 
exponential of (\ref{zz}) must be put zero, and we obtain:
\begin{eqnarray}
\begin{array}{lrlrlr}
\widetilde{\sigma}_z({\bf r}): &  -2\lambda_{\sigma}\sigma^{(0)}_z+
2\mu_BH(1+\frac 87 \tau^{(0)}_z)=0; &
\widetilde{\sigma}_x({\bf r}): &  \sigma^{(0)}_x=0; &
\widetilde{\sigma}_y({\bf r}): &  \sigma^{(0)}_y=0;\\
\widetilde{\tau}_z({\bf r}): & -2(\lambda_{\tau}+{\cal A}_{\bf 0})\tau^{(0)}_z
+\frac {16}7\mu_BH\sigma^{(0)}_z=0; &
\widetilde{\tau}_x({\bf r}): & {\tau}^{(0)}_x=0; &
\widetilde{\tau}_y({\bf r}): & {\tau}^{(0)}_y=0.
\end{array}
\nonumber
\end{eqnarray}
This yields
$$
\tau^{(0)}_z=\frac 78\frac z{\lambda_{\tau}+{\cal A}_{\bf 0}-z}\,,\quad
\sigma^{(0)}_z=\frac{\mu_BH}{\lambda_{\sigma}}
\frac {\lambda_{\tau}+{\cal A}_{\bf 0}}{\lambda_{\tau}+{\cal A}_{\bf 0}-z}\,,
$$
where $z=(8\mu_BH/7)^2/\lambda_{\sigma}$.

Then, extracting the constants and the contribution of the uniform components 
of $\sigma$- and $\tau$-fields (${\cal Z}_0$), 
we arrive at the Gaussian integration over $\widetilde{\sigma}$ and 
$\widetilde{\tau}$ in ${\cal Z}_1$ (${\cal Z}={\cal Z}_0{\cal Z}_1$):
\begin{eqnarray}
{\cal Z}_1=\prod_{{\bf r}}\!{\displaystyle \int_{-\infty}^{\infty}}\!
d \widetilde{\bbox\sigma}({\bf r})\!
{\displaystyle \int_{-\infty}^{\infty}}\!
d \widetilde{\bbox\tau}({\bf r})
\exp \beta\left\{-\lambda_{\sigma}
\widetilde{\bbox\sigma}^2({\bf r})
+\frac {16}7\mu_BH\widetilde{\sigma}_z({\bf r})\widetilde{\tau_z}({\bf r})
\right.
\nonumber\\
\left.-\lambda_{\tau}\widetilde{\bbox\tau}^2({\bf r})
-\sum_{{\bf r}'}{\cal A}_{\alpha\beta}({\bf r}-{\bf r}')
\widetilde{\tau}_{\alpha}({\bf r})
\widetilde{\tau}_{\beta}({\bf r}')
\right\}\,,
\label{z1}
\end{eqnarray}
where ${\cal Z}_0=\exp (-N{\cal F}_0/T)$ and
\begin{eqnarray}
\label{f0}
-{\cal F}_0= \frac 34\lambda_{\sigma}+\frac 12\lambda_{\tau}+
\left(\frac 78\right)^{\!2}\left(z+
\frac{z^2}{\lambda_{\tau}\!+\!{\cal A}_{\bf 0}-\!z}\right)\,.
\end{eqnarray}
The $\widetilde{\sigma}$ integration can be easily performed. 
It results in the contribution
\begin{eqnarray}
\label{f1s}
-{\cal F}_1^{(\sigma)}=\frac 32 T\ln\frac{\pi T}{\lambda_{\sigma}},
\end{eqnarray}
and transforms ${\cal A}_{zz}({\bf r}-{\bf r}')\to
{\cal A}_{zz}({\bf r}-{\bf r}')-\delta_{{\bf r},{\bf r}'}z$.

Thus, we can rewrite ${\cal Z}_1$ as 
$\exp(-N({\cal F}_1^{(\sigma)}+{\cal F}_1^{(\tau)})/T)$, and
\begin{eqnarray}
\label{z2}
\exp -\frac {N{\cal F}_1^{(\tau)}}T=
\prod_{{\bf r}}\!{\displaystyle \int_{-\infty}^{\infty}}\!
d \widetilde{\bbox\tau}({\bf r})\exp -\beta
\sum_{{\bf r}'}\widetilde{{\cal A}}_{\alpha\beta}({\bf r}-{\bf r}')
\widetilde{\tau}_\alpha({\bf r})\widetilde{\tau}_\beta({\bf r}'),
\end{eqnarray}
where 
\begin{eqnarray}
\widetilde{{\cal A}}_{zz}({\bf r})=\delta_{{\bf r},0}(\lambda_{\tau}-z)
+{\cal A}_{zz}({\bf r}),\quad
\widetilde{{\cal A}}_{xx}({\bf r})=\delta_{{\bf r},0}\lambda_{\tau}+
{\cal A}_{xx}({\bf r}),\quad
\widetilde{{\cal A}}_{xz}({\bf r})={\cal A}_{xz}({\bf r}).
\nonumber
\end{eqnarray}
The functional integration in (\ref{z2}) can be straightforwardly performed:
\begin{eqnarray}
\label{f1t}
-{\cal F}_1^{(\tau)}=
\frac T2\frac 1N\sum_{{\bf k}}\ln\frac{(\pi T)^2}
{\widetilde{{\cal A}}^{zz}[{\bf k}]
\widetilde{{\cal A}}^{xx}[{\bf k}]-(\widetilde{{\cal A}}^{xz}[{\bf k}])^2}\,.
\end{eqnarray}

\begin{figure}
\epsfxsize=16cm
\epsffile{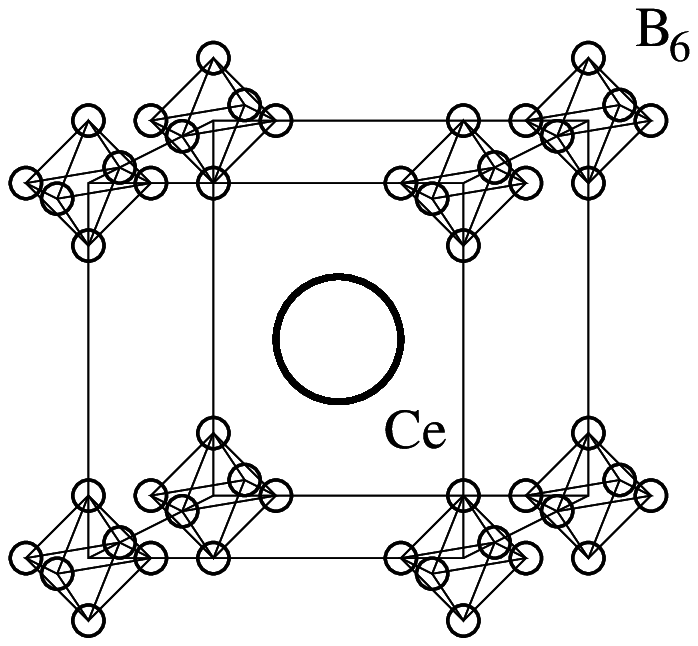}
\hfill
\vspace{0.8cm}

\noindent
\baselineskip=10pt
{{\small FIG.\ 1\quad 
The elementary cubic cell of CeB$_6$. Ce ions, as well as boron octahedrals, 
form simple cubic sublattices with a lattice parameter $a=4.14$ \AA.}}
\end{figure}
\newpage
\begin{figure}
\epsfxsize=19cm
\epsffile{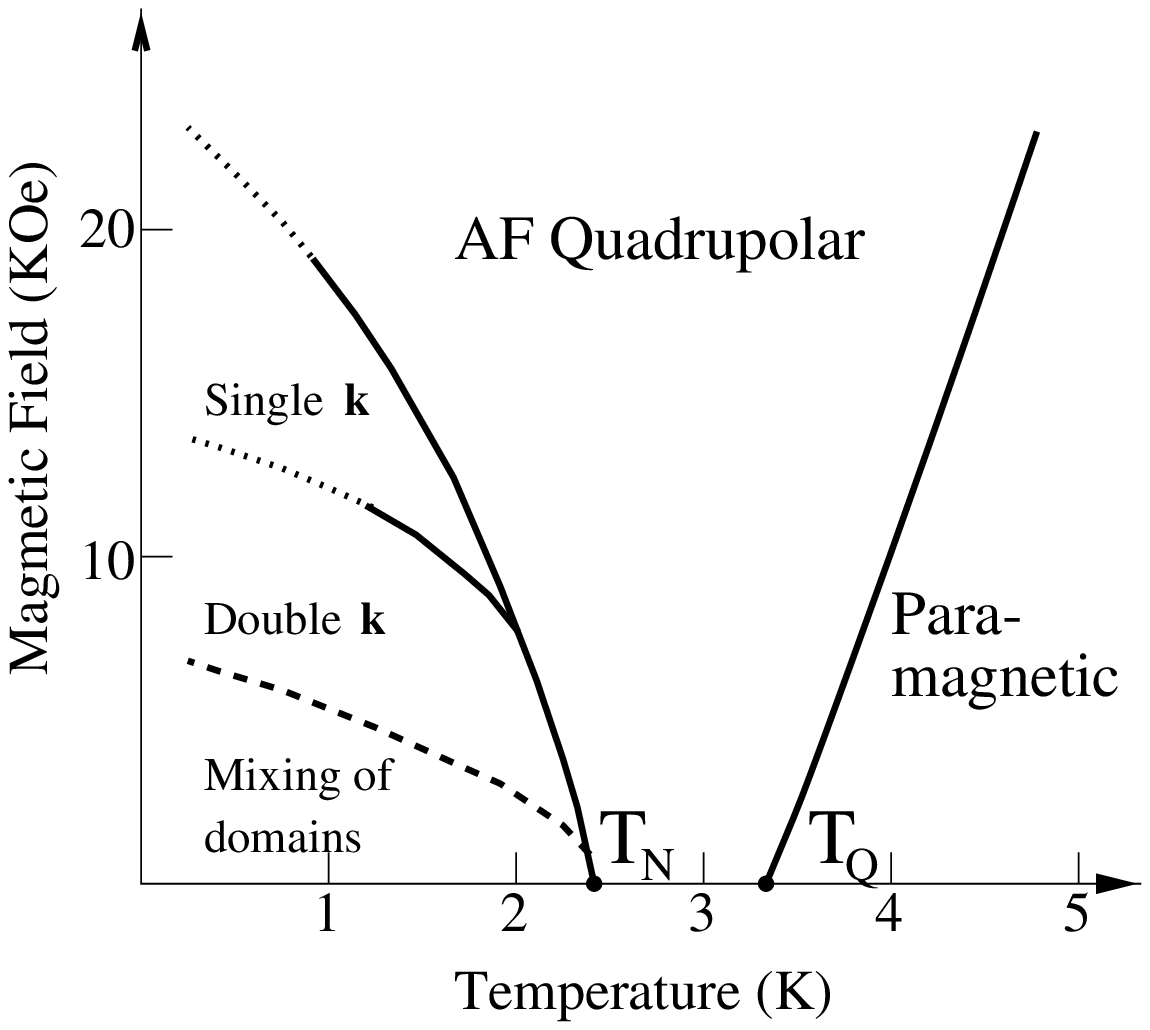}
\hfill
\vspace{0.8cm}

\noindent
\baselineskip=10pt
{{\small FIG.\ 2\quad 
The low field part of the phase diagram. Positions of the lines, confining the 
magnetically ordered phases, depend on the orientation of magnetic field.
}}
\end{figure}
\newpage
\begin{figure}
\epsfxsize=20.5cm
\epsffile{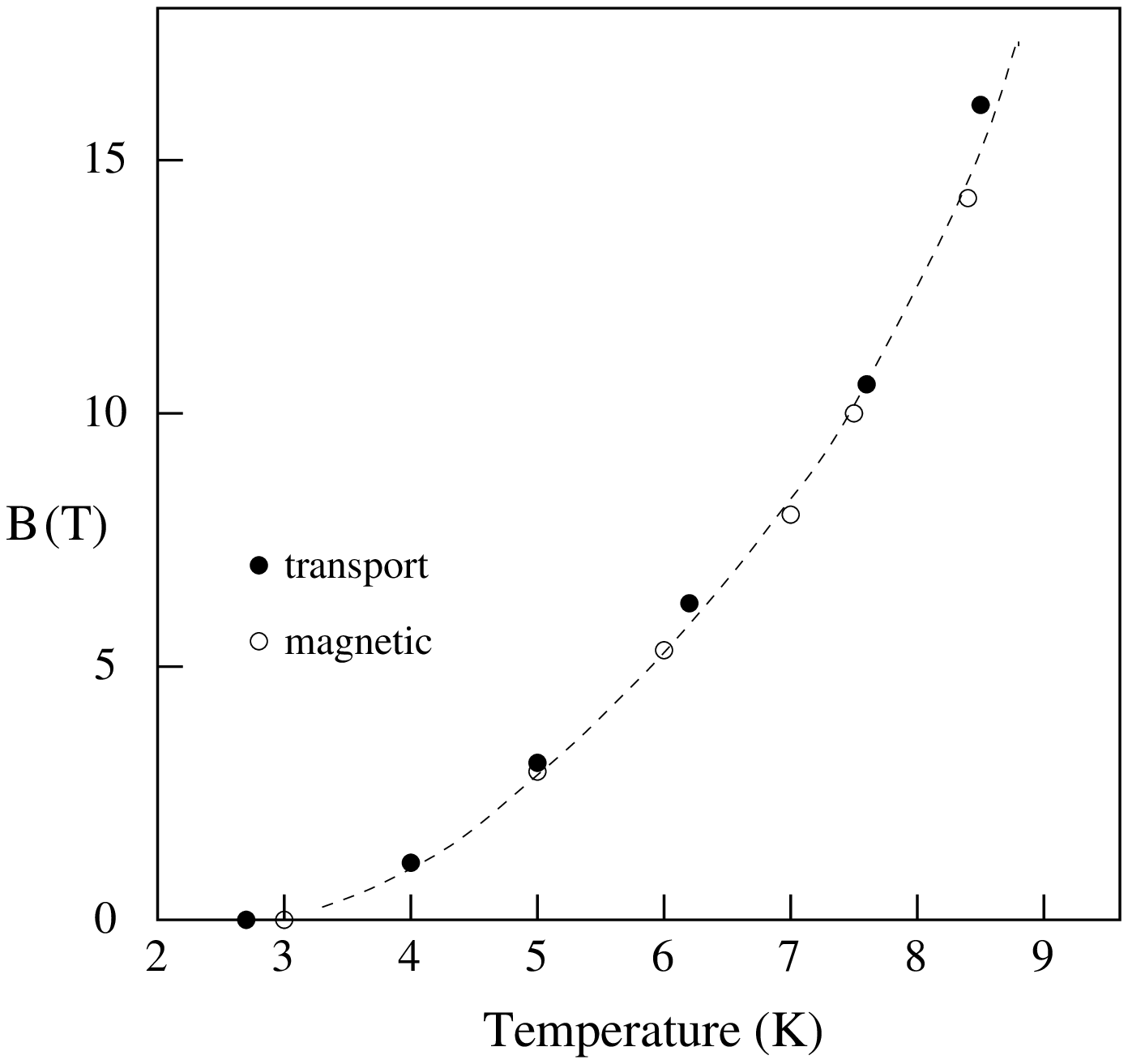}
\hfill

\noindent
\baselineskip=10pt
{{\small FIG.\ 3\quad 
The boundary between two phases, AFQ and D, as determind experimentally 
\cite{lacerda} by transport (magnetoresistance) and magnetic measurements,
full and open circles, respectively. A line is included as a guide to the eye.
}}
\end{figure}
\newpage
\begin{figure}
\epsfxsize=20.5cm
\epsffile{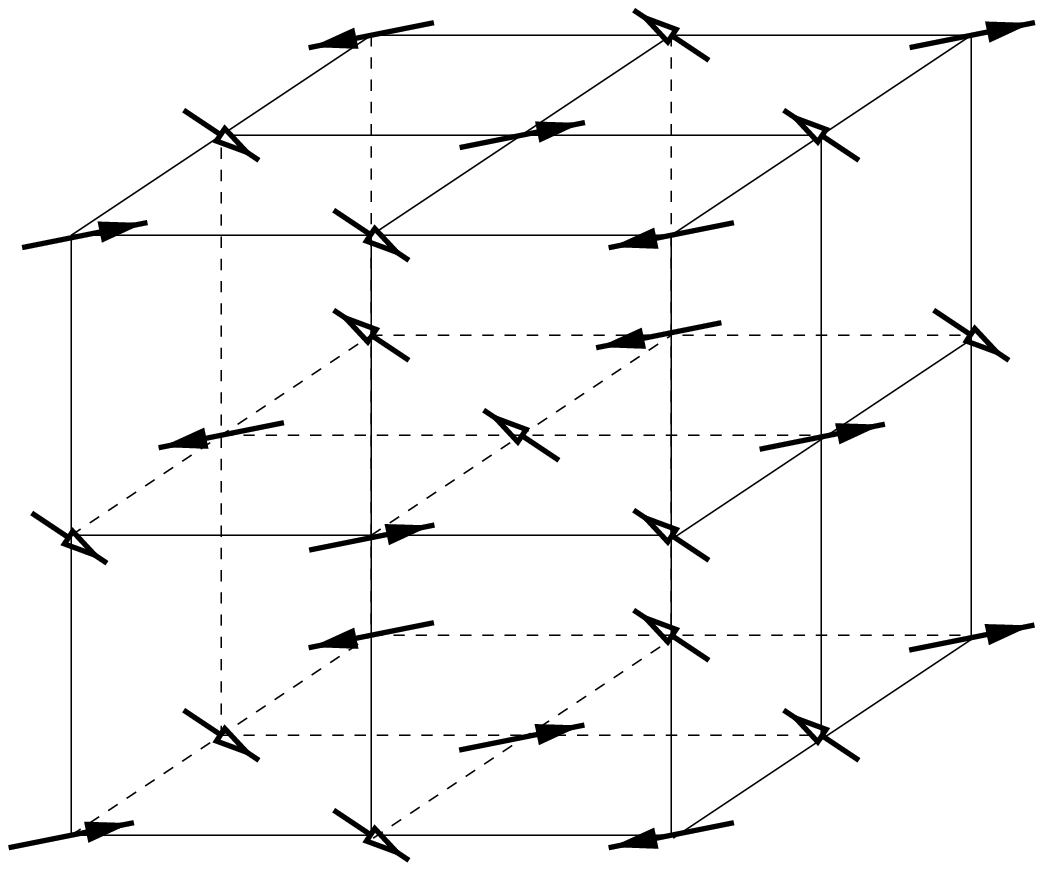}
\hfill

\noindent
\baselineskip=10pt
{{\small FIG.\ 4\quad 
One of the possible arrangements of magnetic moments in the double-${\bf k}$ 
structure.
}}
\end{figure}
\newpage
\begin{figure}
\epsfxsize=14cm
\epsffile{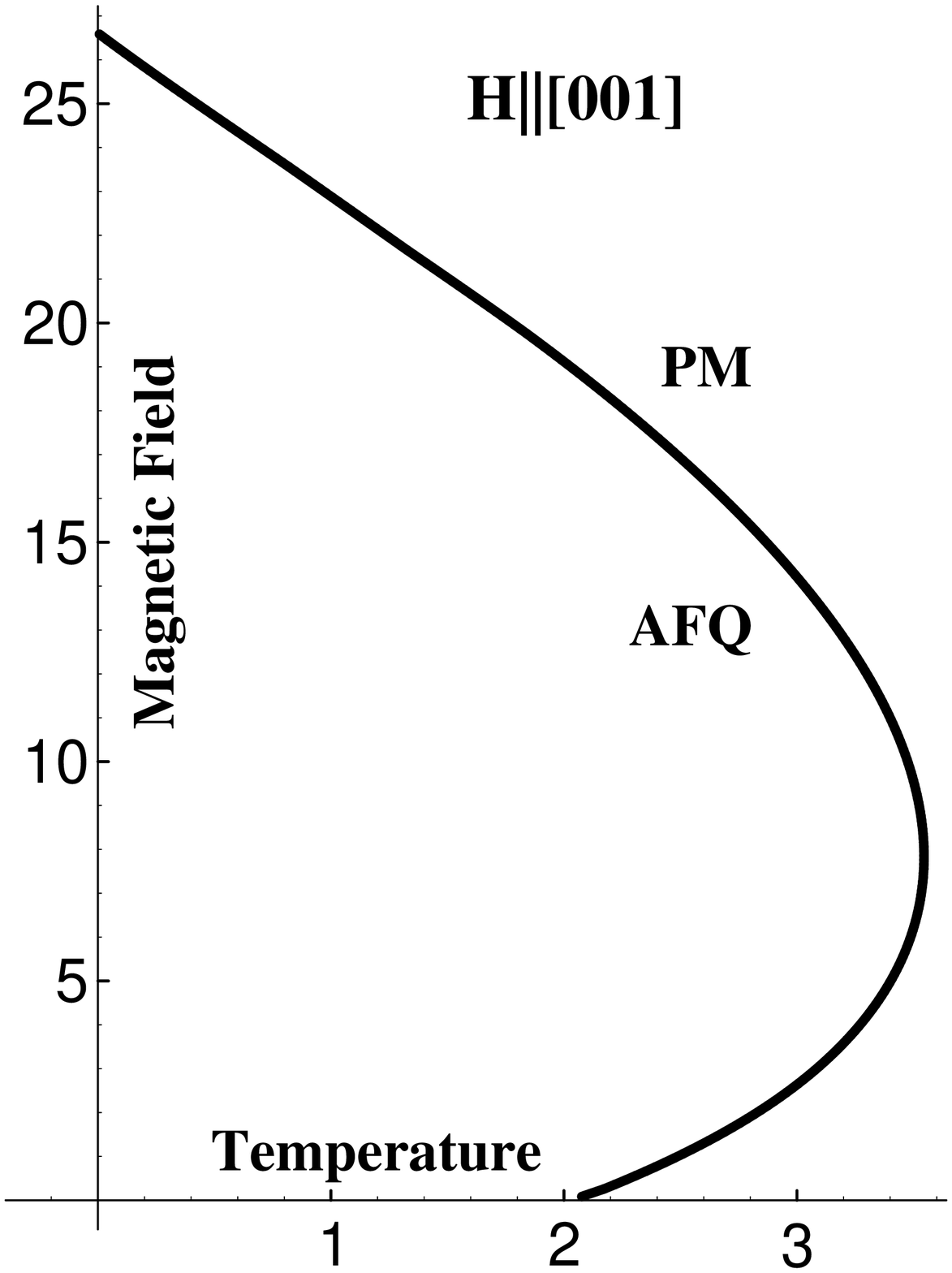}
\hfill

\noindent
\baselineskip=10pt
{{\small FIG.\ 5\quad The line of the AFQ-D phase transition according to 
Eqs.(26)-(27). Units for $T$ and $H$ are discussed in the text.}}
\end{figure}
\newpage
\begin{figure}
\epsfxsize=14cm
\epsffile{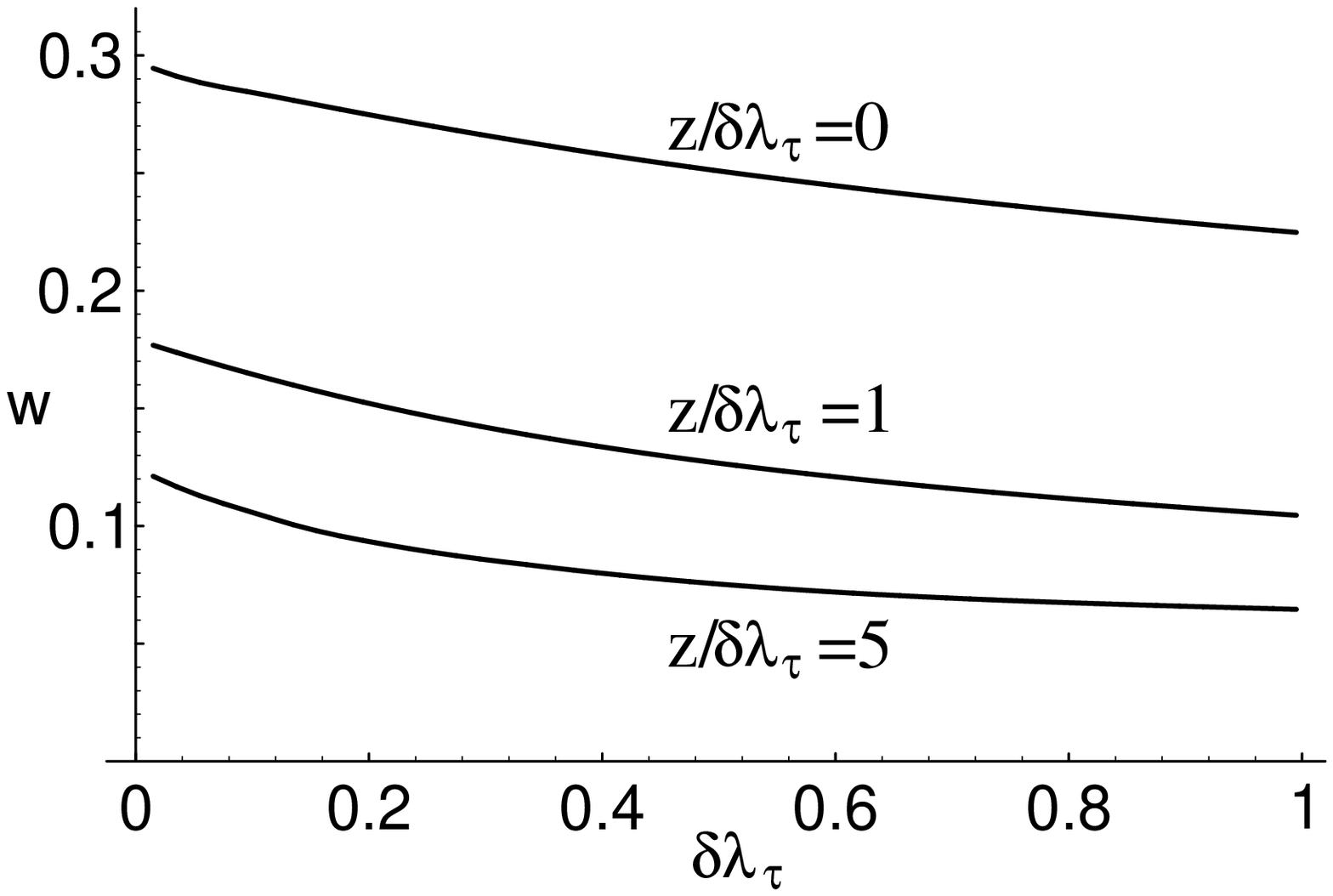}
\hfill
\vspace{-3cm}

\noindent
\baselineskip=10pt
{{\small FIG.\ 6\quad $w(z/\delta\lambda_\tau;\delta\lambda_\tau)$ as function 
of $\delta\lambda_\tau$ (Eq.(\ref{w_f})).
}}
\end{figure}
\newpage
\begin{figure}
\epsfxsize=16cm
\epsffile{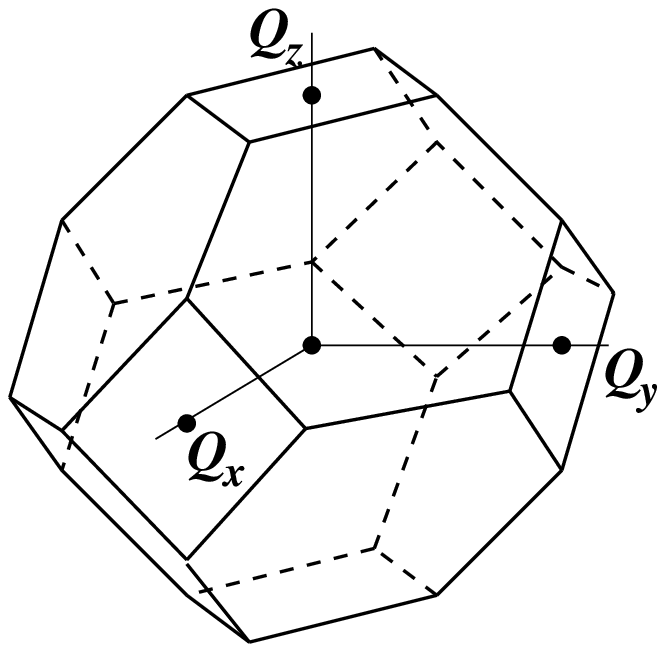}
\hfill
\vspace{0.8cm}

\noindent
\label{br_zone}
\baselineskip=10pt
{{\small FIG.\ 7\quad The first Brillouin zone of reciprocal space of the fcc structure.}}
\end{figure}
\newpage
\begin{figure}
\epsfxsize=22cm
\epsffile{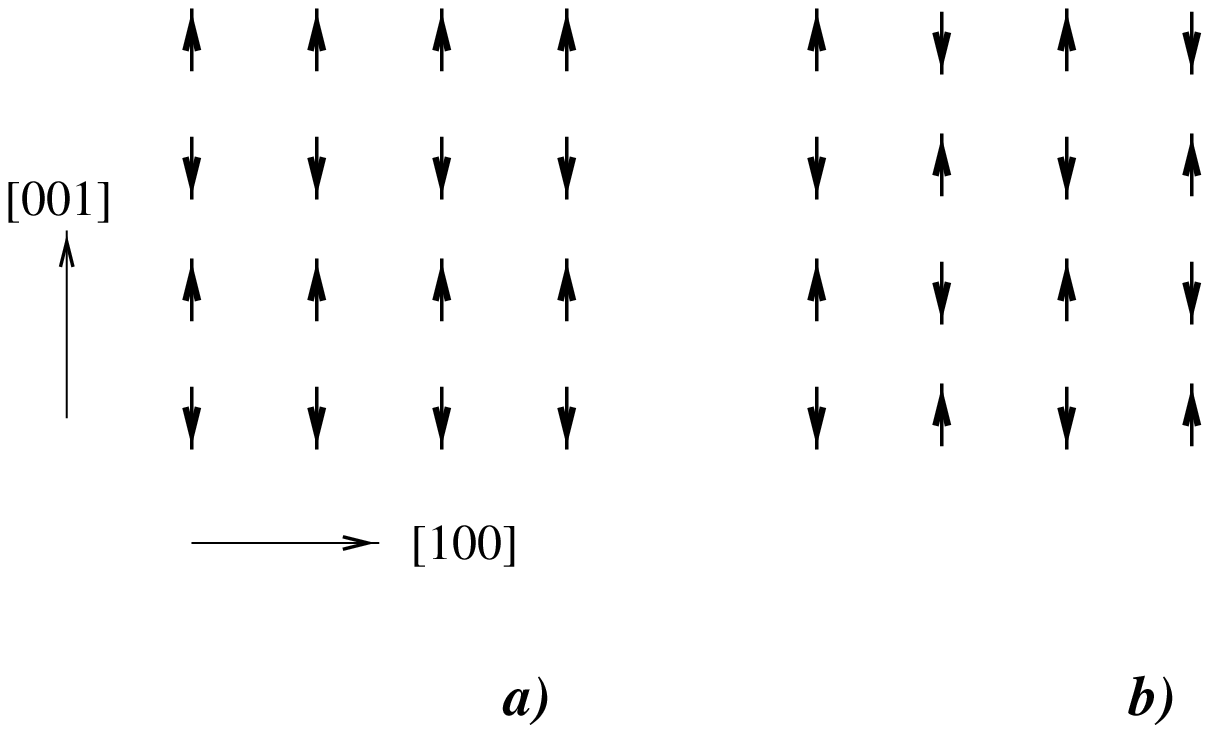}
\hfill
\vspace{3cm}

\noindent
\baselineskip=10pt
{{\small FIG.\ 8\quad 
Arrangement of staggered magnetization for magnetic field applied
along [001]; (a) NMR interpretation [11], (b) ${\bf Q}$-modulated
structure.}}
\end{figure}
\newpage
\begin{figure}
\epsfxsize=23cm
\epsffile{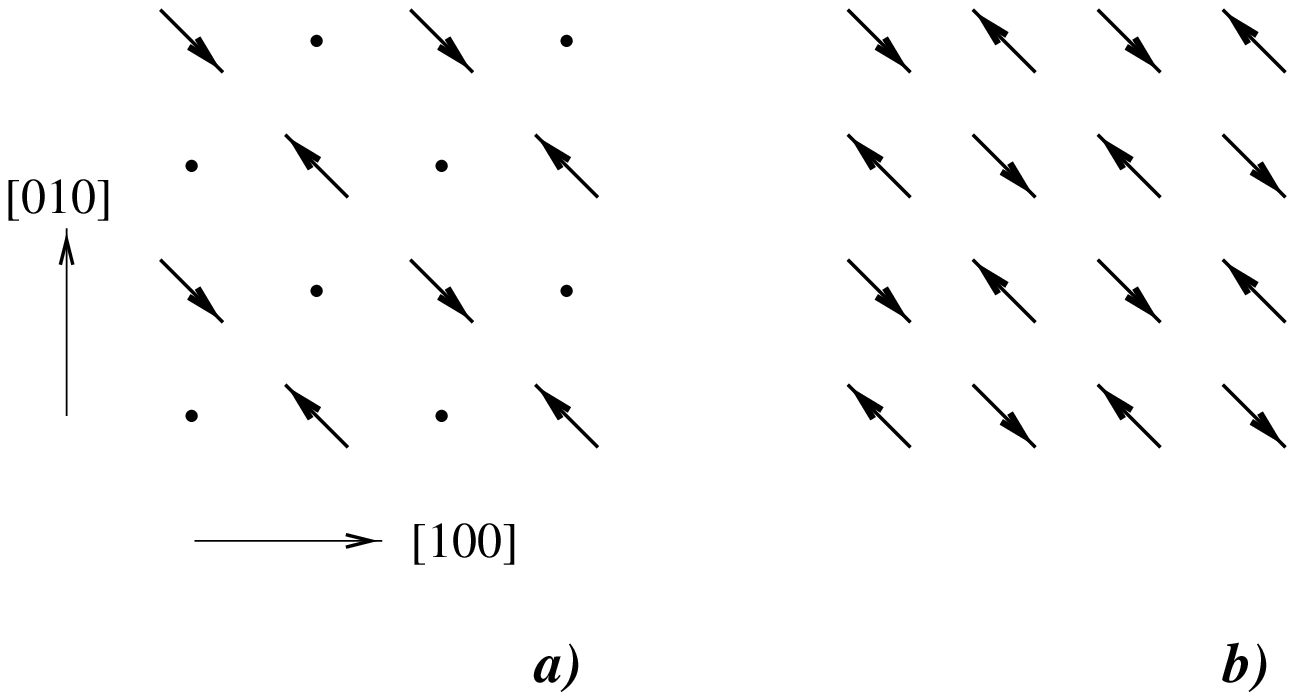}
\hfill
\vspace{3cm}

\noindent
\baselineskip=10pt
{{\small FIG.\ 9\quad 
Arrangement of staggered magnetization for magnetic field applied
along [110]; (a) NMR interpretation [11], (b) ${\bf Q}$-modulated
structure.}}
\end{figure}
\begin{table}[t]
\begin{tabular} {|c||c|c|c|c|}
${\bf k}$ & ${\cal A}_{\bf k}^z$ & ${\cal A}_{\bf k}^x$ 
& ${\cal B}_{\bf k}^x$ & ${\cal B}_{\bf k}^z$ \\ \hline
$\frac 12 \frac 12 \frac 12 $ & $ -10.736 $ & $ -10.736 $ & $ 1.789 $ & $ 1.789 $\\ 
$\frac 12 \frac 12 0 $ & $ -10.478  $ & $ 3.484 $ & $ -0.581 $ & $ 2.909 $\\
$0 0 \frac 12 $ & $ 2.139 $ & $ 7.349 $ & $ -0.357 $ & $ -1.659 $\\
$0 0 0 $ & $ 9.325 $ & $ 9.325 $ & $ -1.554 $ & $ -1.554 $\\ 
\end{tabular}
\vspace{5mm}

\caption[Table I. ]{Strength of the quadrupolar interaction 
at high symmetry points}
\end{table}
\begin{table}
\begin{tabular} {|c||l|l|l|l|}
 & $|1,+\rangle$ & $|1,-\rangle$ & $|2,+\rangle$ & $|2,-\rangle$
\\ \hline\hline
$|1,+\rangle$ & $(\frac 12+\tau_z)(\frac 12+\sigma_z)$ & 
$(\frac 12+\tau_z)\sigma_-$ & $\tau_-(\frac 12+\sigma_z)$ & $\tau_- \sigma_-$ 
\\ \hline
$|1,-\rangle$ & $(\frac 12+\tau_z)\sigma_+$ & 
$(\frac 12+\tau_z)(\frac 12-\sigma_z)$ & $\tau_- \sigma_+ $ & 
$\tau_- (\frac 12-\sigma_z)$\\ \hline
$|2,+\rangle$ & $\tau_+(\frac 12+\sigma_z)$ & $\tau_+\sigma_- $ & 
$(\frac 12-\tau_z)(\frac 12+\sigma_z) $ & $(\frac 12-\tau_z)\sigma_- $\\ \hline
$|2,-\rangle$ & $\tau_+\sigma_+ $ & $\tau_+(\frac 12-\sigma_z) $ & 
$(\frac 12-\tau_z)\sigma_+$ & $(\frac 12-\tau_z)(\frac 12-\sigma_z)$
\end{tabular}
\vspace{5mm}

\caption[Table II. ]{The operator forms for transformations
$|\ell,\sigma\rangle\to|\ell',\sigma'\rangle$.}
\end{table}

\begin{references} 
\bibitem[*]{gu} On leave from Landau Institute for Theoretical Physics,
Chernogolovka, Moscow District 142432, Russia.
\bibitem{winzer} K. Winzer, and W. Flesch, J. Physique, {\bf 39}, C6-832 
(1978).
\bibitem{takase} A. Takase, K. Kojima, T. Komatsubara, and T. Kasuya, 
Solid State Commun., {\bf 36}, 461 (1980).
\bibitem{sato84} N. Sato, S. Kunii, I. Oguro, T. Komatsubara, and T. Kasuya,
J. Phys. Soc. Japan {\bf 53}, 3967 (1984).
\bibitem{zirn} E. Zirngiebl, B. Hillebrands, S. {Blumenr\"o\-d\-er}, 
G. {G\"u\-n\-therodt}, M. Loewenhaupt, J. {Ca\-r\-pen\-ter}, K. Winzer, and 
Z. Fisk, Phys. Rev. {\bf 30}, 4052 (1984).
\bibitem{burlet} P. Burlet, J.X. Boucherle, J. Rossat-Mignod, J.W. Cable, W.C. 
Koehler, S. Kunii, and T. Kasuya, J. Physique {\bf 43}, C7-273 (1982).
\bibitem{ohkawa} F.J. Ohkawa,  J. Phys. Soc. Jpn. {\bf 52}, 3897 (1983).
\bibitem{lee} K.N. Lee, and B. Bell, Phys. Rev. {\bf B 6}, 1032 (1972).
\bibitem{fujita} T. Fujita, M. Suzuki, T. Komatsubara, S. Kunii, T. Kasuya and
T. Ohtsuka, Solid State Commun. {\bf 35}, 569 (1980)
\bibitem{peysson} Y. Peysson, C. Ayache, J. Rossat-Mignod, S. Kunii, and 
T. Kasuya, J. Physique {\bf 47}, 113 (1986).
\bibitem{kawa} M. Kawakami, K. Mizuno, S. Kunii, T. Kasuya, H. Enokiya, and 
K. Kume, J. Magn. Magn. Mater. {\bf 30}, 201 (1982).
\bibitem{takigawa} M. Takigawa, H. Yasuoka, T. Tanaka, and Y. Ishizawa,
 J. Phys. Soc. Jpn. {\bf 52}, 728 (1983).
\bibitem{horn} S. Horn, F. Steglich, M. Loewenhaupt, H. Scheuer, W. Flesch, and
K. Winzer, Z. Phys. {\bf B 42}, 125 (1981).
\bibitem{effantin} J.M. Effantin, J. Rossat-Mignod, P. Burlet, H. Bartholin,
S. Kunii, and T. Kasuya, J. Magn. Magn. Mater. {\bf 47}-{\bf 48}, 145 (1985).
\bibitem{uim1} G. Uimin, Y. Kuramoto, and N. Fukushima, Solid State Commun. 
{\bf 97}, 595 (1996).
\bibitem{erk} W.A.C. Erkelens, L.P. Regnault, P. Burlet, J. Rossat-Mignod, 
S. Kunii, and T. Kasuya, J. Magn. Magn. Mater. {\bf 63}-{\bf 64}, 61 (1987).
\bibitem{lacerda} A. Lacerda and M. Torikachvili, unpublished.
\bibitem{feyer} R. Feyerherm, A. Amato, F.N. Gygax, A. Schenck, Y. \={O}nuki,
and N. Sato, Physica {\bf B 194}--{\bf 196}, 357 (1994); 
J. Magn. Magn. Mater. {\bf 140}-{\bf 144} 1175 (1995).
\bibitem{bl} B. Bleaney, Proc. Phys. Soc. (London) {\bf 77}, 113 (1961).
\bibitem{birg} R.J. Birgeneau, M.T. Hutchings, J.M. Baker, and J.D. Riley,
J. Appl. Phys. {\bf 40}, 1070 (1969).
\bibitem{mor} P.M. Levy, P. Morin, and D. Schmitt, Phys. Rev. Lett. {\bf 42},
1417 (1979). 
\bibitem{mor_2} 
According to \cite{mor} the value of the direct quadrupolar 
interaction in some intermetallic rare-earth compounds is of order 
10$^{-3}$ K. That estimate should be consistent with a large screening effect
on a quadrupolar moment by outer electrons, which is unlikely the case of 
CeB$_6$.
\bibitem{Hubbard} J. Hubbard, J.Phys.C {\bf 4}, 53 (1971).
\bibitem{sugi} K. Sugiyama, F. Iga, M. Kasaya, T. Kasuya, and M. Date,
J. Phys. Soc. Jpn. {\bf 57}, 3946 (1988).
\bibitem{matsum} T. Matsumura {\it et al.}, unpublished. 
\bibitem{thes} For a detailed description of orientations and periodicities 
of such domains we may recommend the following thesis: 
J.M. Effantin, Universit\'e de Grenoble (1985); W.A.C. Erkelens, University
of Leiden (1987); A. Bouvet, Universit\'e J. Fourier, Grenoble (1993).
\bibitem{mor2} P. Morin, and D. Schmitt, Phys. Lett. {\bf A 73}, 67 (1979). 
\bibitem{uim2} G. Uimin, Phys. Lett. {\bf A 215}, 97 (1996).
\end{references}
\end{document}